# A scoping review of transfer learning research on medical image analysis using ImageNet

Mohammad Amin Morid, PhD[1], Alireza Borjali, PhD[2,3], Guilherme Del Fiol, MD, PhD[4]

[1]Department of Information Systems and Analytics, Leavey School of Business, Santa Clara University, Santa Clara, CA

[2]Department of Orthopaedic Surgery, Harvard Medical School, Boston, MA

[3]Department of Orthopaedic Surgery, Harris Orthopaedics Laboratory, Massachusetts General Hospital, Boston, MA

[4]Department of Biomedical Informatics, University of Utah, Salt Lake City, UT, USA

**Corresponding Author:**

Mohammad Amin Morid, PhD

Assistant Professor

Department of Information Systems and Analytics

Santa Clara University

500 El Camino Real, Santa Clara, CA, 95053-0382

Phone: +1(408) 554-4629

mmorid@scu.edu

# Abstract

**Objective**: Employing transfer learning (TL) with convolutional neural networks (CNNs), well-trained on non-medical ImageNet dataset, has shown promising results for medical image analysis in recent years. We aimed to conduct a scoping review to identify these studies and summarize their characteristics in terms of the problem description, input, methodology, and outcome.

**Materials and Methods**: To identify relevant studies, MEDLINE, IEEE, and ACM digital library were searched for studies published between June $1^{st}$, 2012 and January $2^{nd}$, 2020. Two investigators independently reviewed articles to determine eligibility and to extract data according to a study protocol defined a priori.

**Results**: After screening of 8,421 articles, 102 met the inclusion criteria. Of 22 anatomical areas, eye (18%), breast (14%), and brain (12%) were the most commonly studied. Data augmentation was performed in 72% of fine-tuning TL studies versus 15% of the feature-extracting TL studies. Inception models were the most commonly used in breast related studies (50%), while VGGNet was the common in eye (44%), skin (50%) and tooth (57%) studies. AlexNet for brain (42%) and DenseNet for lung studies (38%) were the most frequently used models. Inception models were the most frequently used for studies that analyzed ultrasound (55%), endoscopy (57%), and skeletal system X-rays (57%). VGGNet was the most common for fundus (42%) and optical coherence tomography images (50%). AlexNet was the most frequent model for brain MRIs (36%) and breast X-Rays (50%). 35% of the studies compared their model with other well-trained CNN models and 33% of them provided visualization for interpretation.

**Discussion**: This study identified the most prevalent tracks of implementation in the literature for data preparation, methodology selection and output evaluation for various medical image analysis tasks. Also, we identified several critical research gaps existing in the TL studies on medical image analysis. The findings of this scoping review can be used in future TL studies to guide the selection of appropriate research approaches, as well as identify research gaps and opportunities for innovation.

## 1. Introduction

While convolutional neural networks (CNN) were initially explored in computer vision in the 1980s [1], it was not until 2012 that the ImageNet competition demonstrated the potential of using CNN for image analysis. Since then, CNN has become a popular machine learning approach for various applications including medical image analysis.

Full training of a CNN from scratch has two main requirements: 1) a large labeled dataset, and 2) extensive computational and memory resources. In clinical practice, such large labeled datasets are not always available. Creating a large labeled dataset is labor intensive and the number of patients with a specific medical condition of interest might not be sufficient to create a large dataset [2].

An alternative approach to full training of CNN is transfer learning (TL). By leveraging TL, the knowledge gained from large non-medical data can be transferred to solve a targeted medical problem. More specifically, parameters of well-trained CNN models on non-medical ImageNet data with natural images (e.g., AlexNet[3], VGGNet[4] and ResNet[5]) can be transferred to a targeted CNN model to solve a medical imaging problem.

Previous literature reviews focused on the usage of non-TL based deep learning methods [6,7] and TL-based general machine learning methods for medical imaging [8]. Yet, previous reviews have not focused on TL-based deep learning methods from non-medical data (i.e., ImageNet) for medical image analysis. Employing CNN models well-trained on non-medical ImageNet data for medical image analysis is a recent emerging trend; a review on medical imaging analysis up to early 2017 [7] could not find more than a few TL studies on ImageNet. Therefore, this scoping review aimed to summarize medical image analysis studies that used TL approaches on ImageNet. Specifically, we extracted study characteristics such as input data (e.g., dataset size), CNN model, transferring knowledge (i.e., parameters), and performance measures. We aimed to answer the following research questions: 1) What medical image analysis tasks can benefit from using TL on ImageNet data? 2) What are the characteristics of the input data? 3) What TL process (e.g., in terms of the CNN models or transferred parameters) has been followed? 4) What are the outcomes (e.g., performance accuracy)?

## 2. Background

### 2.1. Convolutional Neural Networks

CNN is a machine learning method commonly used in machine vision and medical image analysis [9]. A CNN typically consists of an input layer, one to many convolution layers, pooling operations (or layers), and a fully connected layer [10]. More details about CNNs can be found in [11].

*2.1.1. ImageNet*

The ImageNet Large Scale Visual Recognition Challenge (ILSVRC) is a large scale object recognition challenge, which has been running annually since 2010 [12]. One of the datasets used for this challenge is the ImageNet dataset [13], which contains over 15 million labeled images. Some CNN models have been very successful in classifying images in the ImageNet dataset into its corresponding categories. These models are briefly explained in the following subsections. A more comprehensive description of each model can be found elsewhere [14]. As recently reported in a review by Cheplygina et al., ImageNet is the most commonly used dataset for TL based medical image analysis [8].

*2.1.2. AlexNet*

This CNN model was the winner of ILSVRC2012 [3]. The architecture consists of eight layers. The first layers are convolutional layers followed by a max-pooling layer for data dimension reduction. Rectified linear unit (ReLu) is used for the activation function, which has a fast raining advantage over other activation functions [15]. The remaining three layers are fully connected layers [33].

*2.1.3. VGGNet*

The Visual Geometry Group (VGG) first introduced VGG-16 in ILSVRC2014 followed by VGG-19 as two successful architectures on ImageNet [16]. These models make an improvement over AlexNet by replacing large kernel-sized filters with multiple small kernel-sized filters resulting in 13 and 16 convolution layers for VGG-16 and VGG-19 respectively.

*2.1.4. CaffeNet*

This CNN model is a slight variation of AlexNet. Unlike AlexNet, CaffeNet does not use data augmentation (section B.3) and places the pooling layer before normalization operation. As a result, CaffeNet slightly improves the computational efficiency of AlexNet, since the data dimension reduction happens before normalization operation [17].

*2.1.5. ZFNet*

This CNN model was the winner of ILSVRC2013 and is an improved version of AlexNet with similar eight layers architecture [18]. ZFNet introduced the concept of deconvolutional network [19] to tackle the black-box nature of CNN models by showing how CNN learns feature representations. Deconvolutional network maps the learned features into input pixel space, which improves the CNN interpretability.

*2.1.6. Inception*

GoogLeNet model (also called Inception-V1) attempted at improving the efficiency of VGGNet in terms of memory usage and runtime without reducing accuracy [20]. To achieve this, it eliminated the activation functions of VGGNet that are redundant or zero because of the correlations among them. Therefore, GoogLeNet introduced and added a module called Inception that approximates sparse connections between the activation functions. After Inception-V1 the architecture was further refined in three subsequent versions. Inception-V2 used batch normalization for training [21]. Inception-V3 proposed a factorization method to improve the computational complexity of convolution layers [22]. Inception V-4 introduced a uniform simplified version of the Inception-V3 architecture with more inception modules [23].

*2.1.7. ResNet*

Adding more layers to CNN models can lead to accuracy saturation and vanishing gradients. Residual Learning, which is the backbone of ResNet CNN, aims at solving this problem [24]. CNN models prior to ResNet learned features at different abstraction levels at the end of each convolution layer. Rather than learning features, ResNet learns residuals, which is the subtraction of learned features from input for each convolution layer. This is done by using a concept called identity shortcut connections (i.e., connecting the input of a layer to x layers after that) [5]. Variations of ResNet use a different number of layers, such as ResNet-34, ResNet-50, and ResNet-101.

*2.1.8. Inception-Residual Network*

This CNN model combines the strengths of the Inception and ResNet architectures. As mentioned, Inception effectively learns features at different resolutions within the same convolution layer, while ResNet enables the network to have deeper CNN to learn features that are more complex without losing performance. Inception-Residual Networks combine these strengths in two versions: Inception-ResNet-V1 and Inception-ResNet-V2 [23]. Inception-ResNet-V1 is based on Inception-V3 and Inception-ResNet-V2 is based on Inception-V4.

*2.1.9. Xception*

Xception stands for extreme inception and is a modified version of the Inception-V3 [25]. This CNN model uses depth wise separable convolution to involve the spatial dimension and channel dimension of the image separately in the training process. Xception has almost the same number of parameters as InceptionV3 with slightly better performance on ImageNet.

*2.1.10. DenseNet*

In DenseNet [26], each convolution layer receives the output (i.e., feature maps) of all preceding layers as input and passes its own output (i.e., feature maps) to all subsequent layers. Therefore, each layer obtains the collective knowledge of all preceding layers. The resulting CNN model becomes thinner and more compact due to the decreasing number of feature maps. DensNet has several versions such as DenseNet-121, DeneNet-169, and DenseNet-201.

## 2.2. Transfer Learning

The most common issue with training CNN models for medical image analysis (i.e., full training from scratch) is the lack of large labeled datasets. [27]. TL can help address this limitation by transferring the learned parameters (i.e., network weights) of well-trained CNN models on a large dataset (e.g., ImageNet) to solve medical image analysis problems. To achieve this, the convolutional layers of a well-trained CNN model are either fine-tuned or frozen (i.e., used as is), while the fully connected layers are trained from scratch on the medical dataset. The idea behind TL is that although medical datasets are different from non-medical datasets, the low-level features (e.g., straight and curved lines that construct images) are universal to most of the image analysis tasks [28]. Therefore, transferred parameters (i.e., weights) may serve as a powerful set of features, which reduce the need for a large dataset as well as the training time and memory cost. There are two transfer learning approaches: feature-extracting and fine-tuning [28].

*2.2.1. Feature-extracting*

This approach utilizes a well-trained CNN model on a large dataset (e.g., ImageNet) as a feature extractor for the target domain (e.g., medical). More specifically, all convolution layers of the well-trained CNN model are frozen, while fully connected layers are removed. The convolution layers serve as a fixed feature extractor to adapt to a new (medical) task. Extracted features are then fed to a classifier, which can be new fully connected layers or any supervised machine learning method. Finally, only the new classifier is trained during the training process rather than the entire network [8].

*2.2.2. Fine-tuning*

This approach also utilizes a well-trained CNN model on a large dataset (e.g., ImageNet) as the base and replaces the classifier layers with a new classifier. However, in this method convolution layers of the well-trained CNN model are not frozen and their weights can get updated during the training process. This is done by initializing the weight of the convolution layers with the pre-trained weights of the well-trained CNN model while initializing the classifier layers with random weights. In this method, the entire network is trained during the training process [24].

## 2.3. Data Augmentation

Increasing the size of labeled data usually improves the performance of CNN models. Data augmentation is a method for artificial data generation for training by creating variations of the original dataset [29]. For image data this includes a variety of image manipulation methods such as rotation, translation, scaling, and flipping techniques [30].

The most important consideration for data augmentation is memory and computational constraints. There are two commonly used data augmentation approaches: online and offline. Online data augmentation is performed on-the-fly during training, while offline data augmentation generates the data beforehand and stores it in memory. The online approach saves memory, but results in slower training time. The offline approach is faster in training, but consumes a large amount of memory.

## 2.4. Visualization of Convolutional Neural Networks

It is difficult to interpret CNN black-box models and understand their decision-making process. It is useful to crack this process to make sure that the neural network is concentrating on appropriate parts of the image [31]. In addition, this can reveal new domain knowledge. Visualization of the learned features by CNNs is the most common practice to understand and trust the decision making process of these networks [18]. The most commonly used visualization methods are briefly described in this section, while more details could be found elsewhere [32].

*2.4.1. Activation Maximization*

This method aims at visualizing the most preferred inputs of neurons at each convolution layer. These preferred inputs show what features are learned. The learned features in a specific layer are represented by a synthesized input image that would cause maximal activation of a neuron [33].

*2.4.2. Deconvolution*

This method finds the patterns in the input image that activate a specific neuron (i.e., feature map) of a convolution layer. These patterns are reconstructed by mapping the neuron's feature map back to the image pixel space. This process is implemented by a deconvolutional network (DeconvNet) structure, which forward-passes through the original CNN (i.e., inversed computation of a convolution layer) and performs up-sampling (i.e., reversing the down-sampling of a pooling layer) for a given feature map back to the input image [34].

*2.4.3. Class Activation Mapping*

Class Activation Mapping, also known as heatmap, was proposed by Zhou et al. [35]. Heatmap extracted from class activation mapping techniques is a simple method to determine the discriminative image regions used by a CNN model to classify images. This is done by visualizing the trigger of activation functions of intermediate convolution layers [36].

# 3. Method

## 3.1. Overview

Overall, this scoping literature review followed the PRISMA guidelines [37] and the methodological framework for scoping reviews proposed by Arksey et al [38]. Overall, the scoping review had two main goals. First, an analytical goal to identify the most prevalent approaches in the literature for data preparation, methodology selection and output evaluation for various medical image analysis tasks. It should be noted that method prevalence does not imply better efficacy of that method. Finding the most optimal method can only be achieved by benchmarking all methods against each other through direct comparisons using the same dataset. Second, we aimed to identify the research gaps based on the findings from the first goal. Specifically, we aimed to address the following research questions: 1. For what medical tasks ImageNet based models can be effective? Is the prediction task nominal or numerical? 2. What is the image type? What is the required dataset size for achieving a satisfactory performance? Is there any need for data augmentation? 3. What transfer learning approaches are most prevalent? 4. What ImageNet based models are most prevalent? Is there any other classifier that fully connected layers used for the final classification task? 5. What is the best achieved performance in each study? What is the performance of other well-trained CNN models for this specific task? 6. For which problems researchers have been able to provide interpretation using visualization techniques?

## 3.2. Literature search

We searched for eligible articles in MEDLINE, IEEE, and the ACM digital library. Since the ImageNet dataset was initially released in 2012, the results were limited to the studies published after June 1$^{st}$ 2012 up to January 2nd, 2020. The search strategies for each database can be found in Table S1 of the online supplement.

## 3.3. Inclusion and exclusion criteria

We included original research studies focused on classification problems of macroscopic medical images (e.g., X-Rays, Computerized Tomography, Magnetic Resonance Imaging) that directly used CNN models well-trained on non-medical images in ImageNet without any manipulation (i.e., methodological improvement, partial transfer of convolution layers of well-trained CNN models, combination of different models). We excluded studies focusing on microscopic images (e.g., tissue biopsies in Pathology), since the analytical approach for these types of image is very different. We also excluded studies lacking key information for the core study characteristics listed in Table 1. Otherwise, we had no other exclusion criteria such as article format or publication venue.

### 3.4. Study selection

To assess inclusion eligibility, two reviewers independently evaluated the title and abstract of each retrieved article. The same reviewers independently evaluated the full text of potentially eligible studies. Disagreements were resolved through consensus between the two reviewers. The Cohen's kappa interrater agreement was 0.81 for title/abstract screening and 0.86 for full-text screening.

### 3.5. Data extraction

The following 13 features were extracted from the included studies to answer the research questions listed in Table 1 in terms of problem description, input, process (i.e., methodology), and output.

### 3.6. Data analysis

Included studies were summarized according to the characteristics laid out in Table 1. We also provided descriptive statistics in a graphical format to convey the frequency of use of different modeling approaches according to the medical task, anatomical site, image type, data size and augmentation method, transfer learning approach, and visualization method.

**Table 1**: Features extracted from each study.

| Research Question | Category | Feature | Description |
|---|---|---|---|
| 1. For what medical tasks ImageNet based TL models can be effective? Is the prediction task nominal or numerical? | Problem | Medical task | Describes the medical goal of transfer learning |
| | | Anatomical site | Determines the body organ or area involved |
| | | Classification type | Numeric or nominal and if nominal, how many classes |
| 2. What is the image type? What is the required dataset size for achieving a satisfactory performance? Is there any need for data augmentation? | Input | Image type | Imaging modality (e.g., x-ray, MRI, ultrasound) |
| | | Dataset size | Number of cases in the dataset used for training and testing |
| | | Augmentation | Choice of online or offline augmentation and the final size of the dataset used |
| 3. What transfer learning approaches are most prevalent? | Process | Transferred knowledge | Transfer learning approach (i.e., feature-extracting or fine-tuning) |
| 4. What ImageNet based models are most prevalent? Is there any other classifier that fully connected layers used for the final classification task? | | CNN model | The ImageNet based model with the best performance |
| | | Classifier | Whether a fully connected layer or a different classifier is used for classification |
| 5. What is the best achieved performance in each study? What is the performance of other well-trained CNN models for this specific task? | Output | Performance | Highest achieved performance based on the primary outcome |
| | | Benchmark | Models used as a baseline for comparison |
| 6. For which problems researchers have been able to provide interpretation using visualization techniques? | | Visualization | Visualization method used for model interpretation |

## 4. Results

The search resulted in 8,421 studies; after title and abstract review, 689 were selected for full-text and 102 studies met the inclusion criteria described in section 3.2 (Figure 1). Figure 2 shows the distribution of the included studies according to their publication year. Most of the studies (85%) have been published after 2018. A complete list of the included studies and their characteristics is available in the online supplement (Tables S3 to S10). Table S2 contains a list of abbreviations used in the manuscript. Table 2 classifies studies according to CNN model category and image modality.

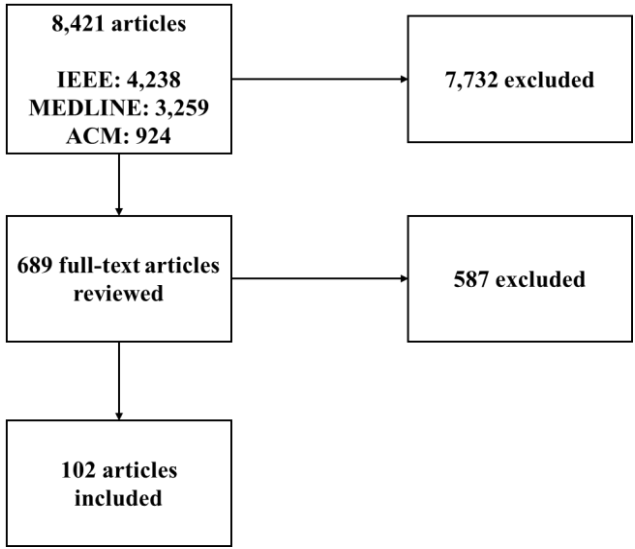

**Figure 1:** Inclusion flow of the scoping review.

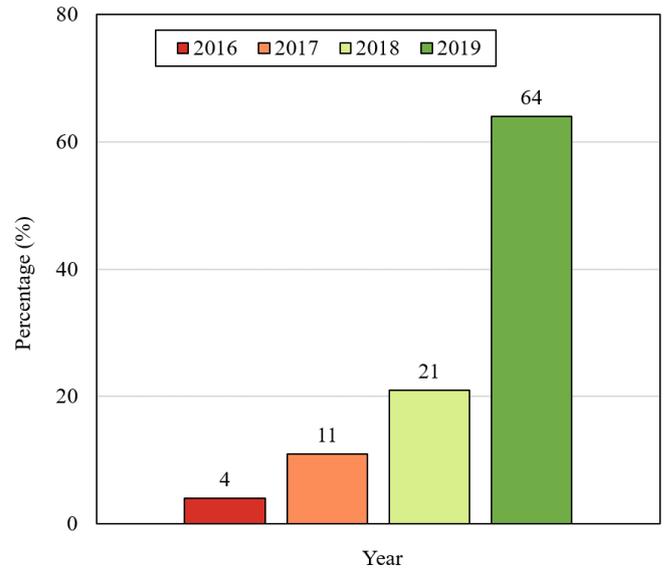

**Figure 2:** Distribution of the included studies according to publication year.

**Table 2**: Distributions of studies over method categories and image types.

|  |  | Image modality | | | | | | | |
|---|---|---|---|---|---|---|---|---|---|
|  |  | **X-ray** | **MRI** | **Fundus** | **Ultrasound** | **CT** | **Endoscopy** | **Skin lesion** | **OCT** |
| **Model Category** | **Inception** | [27,39–46] | [47–51] | [52,53] | [54–59] | [60,61] | [62–65] | [66] | [67] |
|  | **VGGNet** | [68–73] | [74–76] | [77–81] | [82–84] | [85–87] | [88] | [89–91] | [92–94] |
|  | **ResNet** | [95–97] | [98–101] | [102–105] | [106] | [107–109] | [110,111] | [112] | [113] |
|  | **AlexNet** | [114–118] | [119–123] | [124] |  | [9,125,126] |  | [127] |  |
|  | **DenseNet** | [128–133] |  |  | [134] |  |  |  | [135] |
|  | **InceptionResNet** | [136] |  |  | [137] |  |  | [138] |  |

Table 3 shows descriptive statistics of the extracted features (see Table 1 for an explanation of extracted features). X-Ray and magnetic resonance imaging (MRI) were the most commonly used types of images with 29% and 17% frequency respectively. Eye, breast and brain were the most studied organs with 18%, 14% and 12% frequency respectively. The most frequently used CNN models overall, irrespective of the body organ or imaging modality, were Inception-V3 (19%), VGG-16 (18%), AlexNet (15%), and ResNet-50 (13%). Over half of the studies (54%) performed some kind of data augmentation. The majority of studies (65%) did not benchmark their CNN model against any other model. While ILSVRC was a 1000 category classification challenge based on ImageNet, most medical TL studies (71%) performed a binary classification.

**Table 3**: Frequency of study characteristics.

| Feature | Value | Frequency (%) |
|---|---|---|
| **Anatomical site** | Eye | 18 |
| | Breast | 14 |
| | Brain | 12 |
| | Lung | 8 |
| | Skin | 7 |
| | Tooth | 7 |
| | Thyroid | 6 |
| | Stomach | 6 |
| | Others | 24 |
| **Transfer Learning Approach** | Fine tuning weights | 67 |
| | Feature-extracting | 33 |
| **Visualization** | None | 67 |
| | Heatmap | 23 |
| | Deconvolution | 8 |
| | Activation Maximization | 3 |
| **Final Classifier** | Fully connected layer | 84 |
| | Others | 16 |
| **Benchmark** | None | 65 |
| | 1 | 13 |
| | 2 | 11 |
| | >2 | 11 |
| **CNN Model** | Inception | 29 |
| | VGGNet | 26 |
| | ResNet | 19 |
| | AlexNet | 15 |
| | DenseNet | 8 |
| | InceptionResNet | 3 |
| **Image Type** | X-ray | 29 |
| | MRI | 17 |
| | Fundus | 12 |
| | Ultrasound | 12 |
| | CT | 11 |
| | Endoscopy | 7 |
| | Skin lesion | 7 |
| | OCT | 6 |
| **Augmentation** | None | 46 |
| | Offline | 47 |
| | Online | 7 |
| **Classification Task** | Binary | 71 |
| | Categorical | 25 |
| | Numeric | 4 |

Figures 3 shows the frequency of studies using specific types of TL CNN models per image type. Inception models were the most frequently used models for studies that analyzed X-Rays (31%), endoscopic images (57%), and ultrasound images (55%). GoogLeNet and AlexNet (29% each) were the most frequent models for MRIs. VGGNet models were the most commonly used for studies analyzing skin lesions (43%), fundus images (42%) and OCT data (50%). Three CNN models were used with similar frequency in CT scan studies.

Figure 4 shows the frequency of studies using specific types of TL CNN models per anatomical site. Various versions of Inception model were the most frequent approach in studies analyzing breast images (50%), while VGGNet was the most frequent in studies involving eye (44%), skin (50%) and tooth (57%) images. AlexNet and DenseNet were the most frequent model in brain (42%) and lung studies (38%).

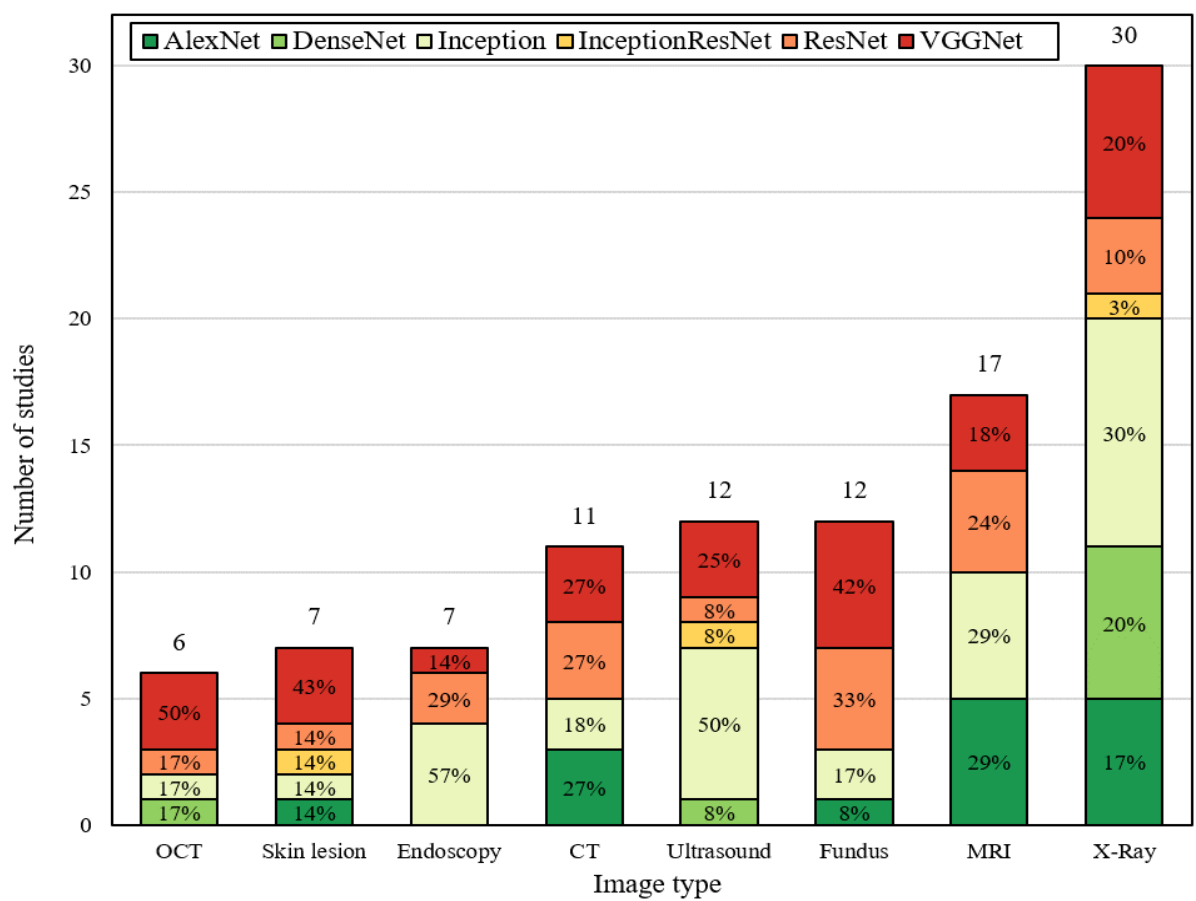

**Figure 3:** Frequency of studies using specific types of TL CNN models per image type.

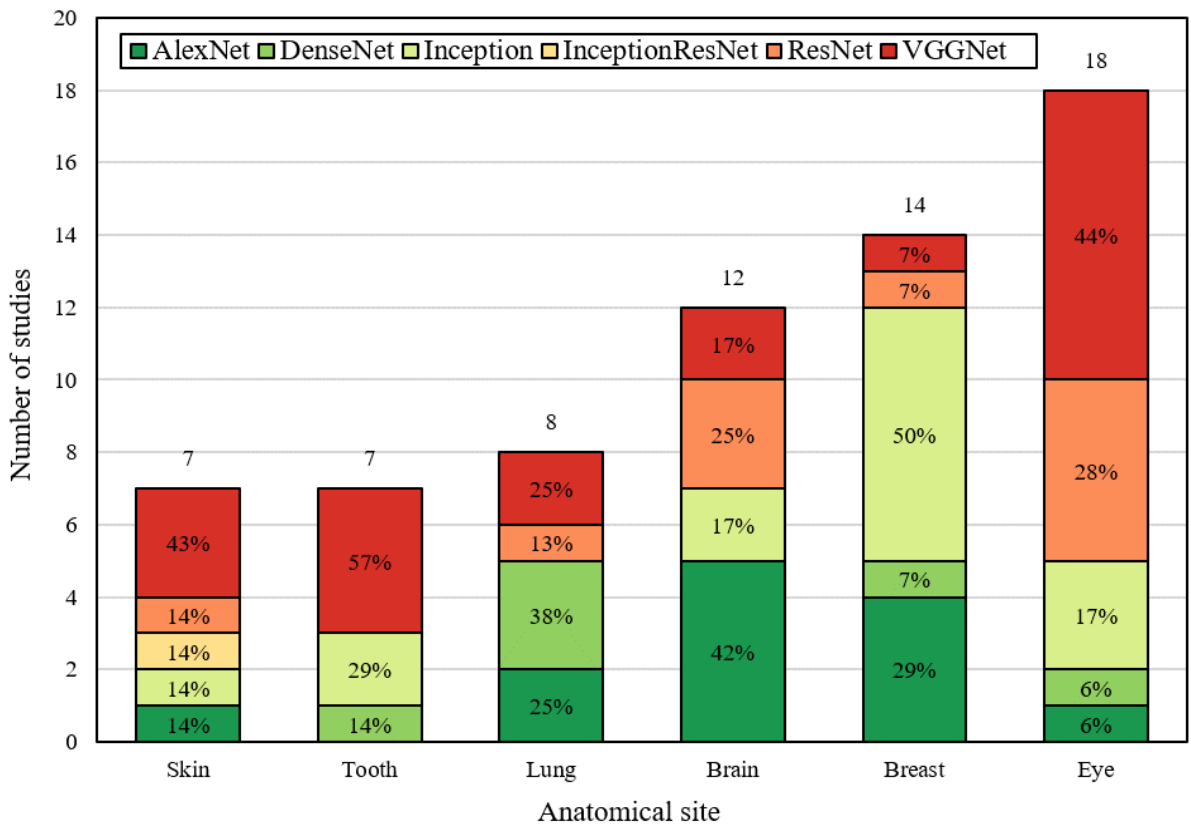

**Figure 4:** Frequency of studies using specific types of TL CNN models per anatomical site. Only anatomical sites with at least 5% overall representation in the included studies are shown.

Figure 5 combines Figure 3 and Figure 4 by considering both imaging modality and anatomical site at the same time. GoogLeNet (combined with SVM classifier) was used in 100% of the studies that analyzed breast MRI, while AlexNet was the most commonly used CNN model (36%) for studies that analyzed brain MRI. Inception models (especially Inception-V3) were the most frequent (57%) among the studies that analyzed skeletal system X-Rays (i.e., hip, knee and wrist). AlexNet (50%), DenseNet (60%) and VGGNet (67%) were the most commonly used models for studies that analyzed breast, lung and tooth X-rays respectively. Only a few studies analyzed CT scans with no predominant CNN model.

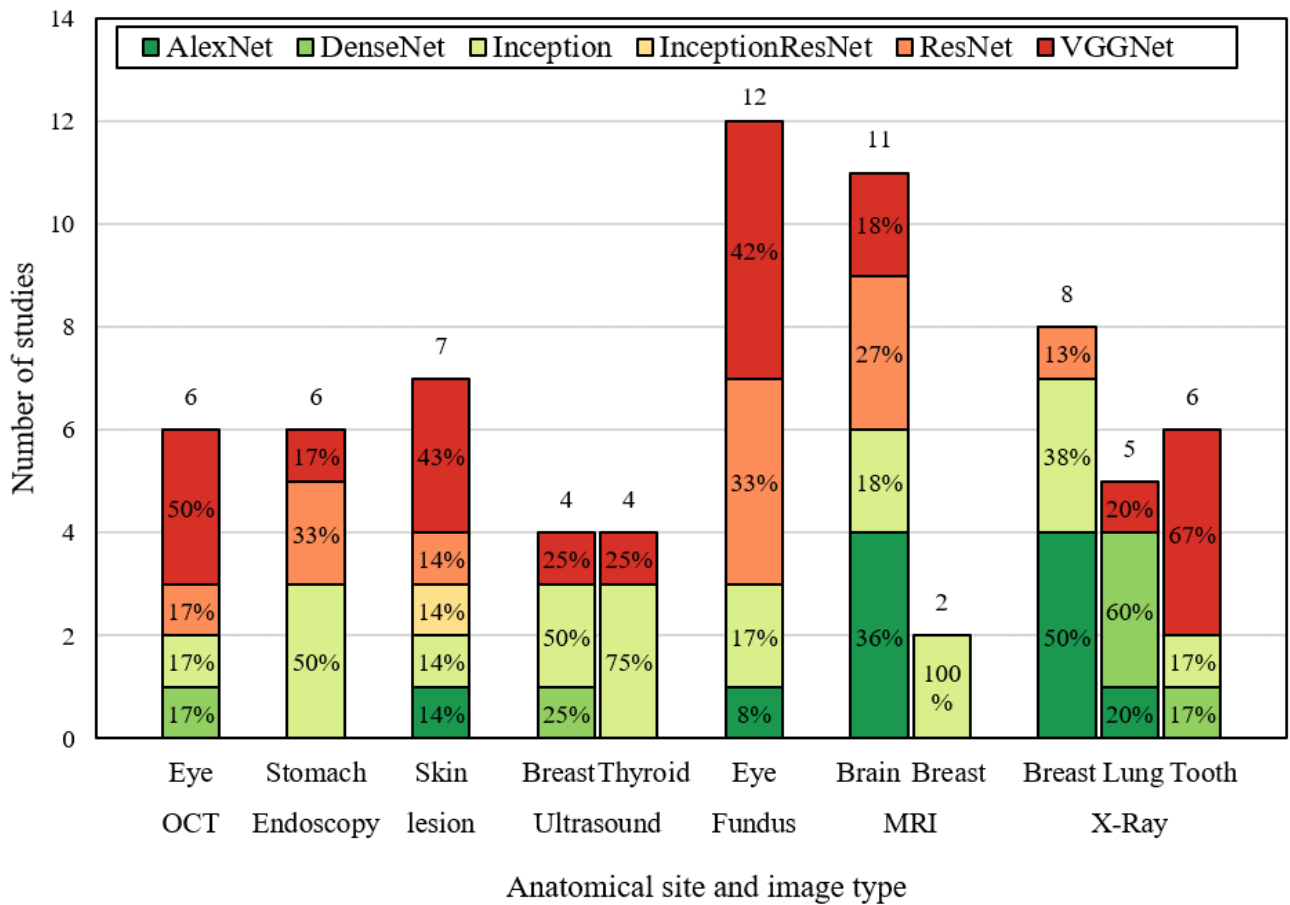

**Figure 5:** Frequency of studies using specific types of TL CNN models per image type and anatomical site. Only anatomical sites and image types with at least 5% overall representation in the included studies are shown.

Figure 6 and Figure 7 show the frequency of transfer learning approaches with and without data augmentation, and per dataset size respectively. Data augmentation was more prevalent among studies that employed fine-tuning TL (72% of fine-tuning TL studies versus 15% of the feature-extracting TL studies). Moreover, among the studies with less than 1,000 images, 22% of the feature extracting TL studies and 77% of fine-tuning TL studies performed data augmentation. Similar patterns were observed among studies with 1,000 to 10,000 images (10% vs. 77%), as well as those with over 10,000 images (0% vs. 55%).

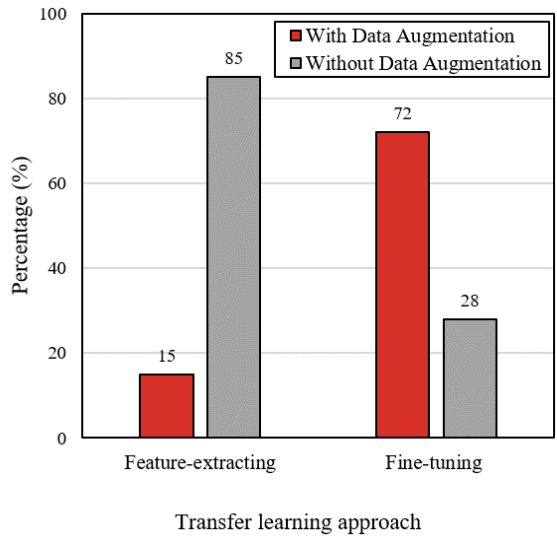
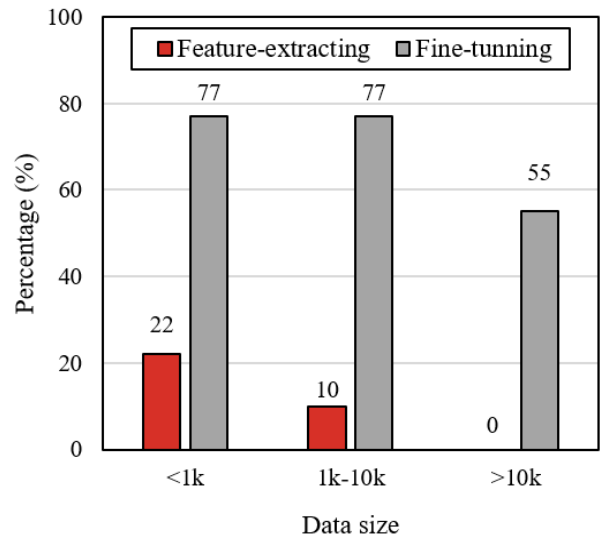

**Figure 6:** Frequency of transfer learning approaches in studies with and without data augmentation.

**Figure 7:** Frequency of transfer learning approaches in studies with data augmentation according to different dataset sizes.

Figure 8 shows the frequency of different visualization methods per anatomical site. 33% of the reviewed studies attempted to provide CNN model visualization, mostly through heat maps (67%) (see Table 3). Studies analyzing images of the brain (58%), lung (50%), and tooth (43%) were the ones to most frequently include a visualization approach.

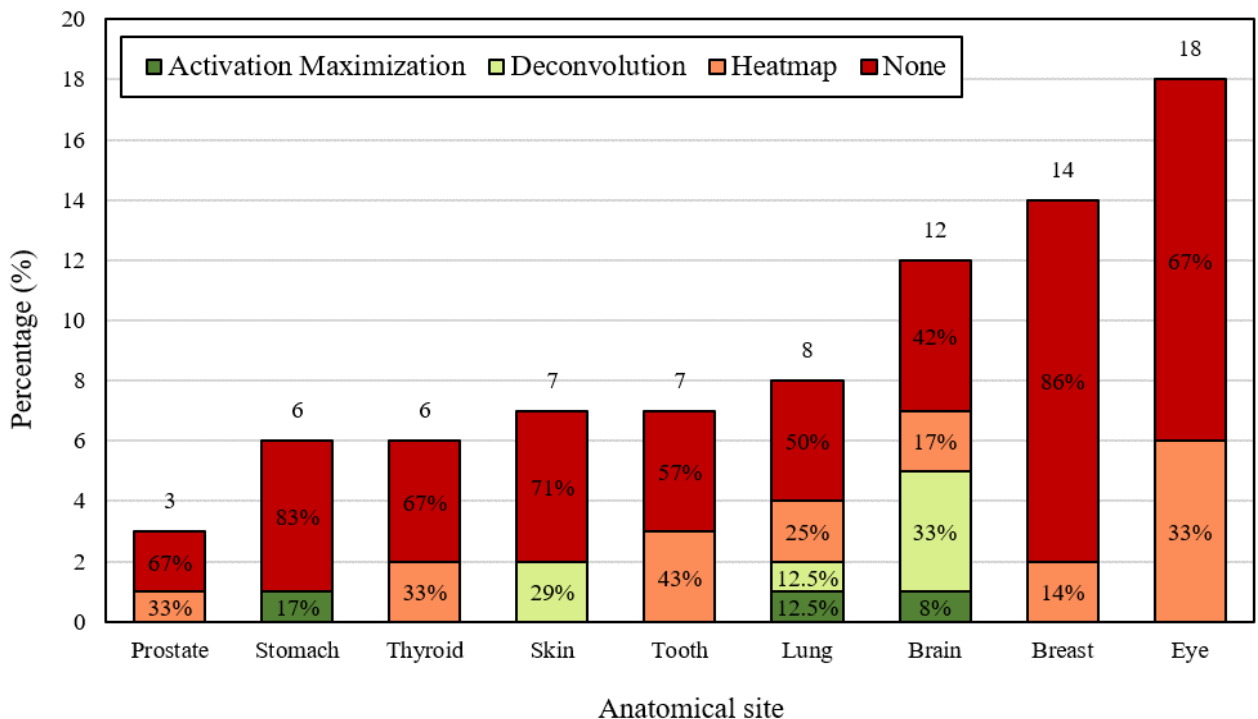

**Figure 8:** Frequency of different visualization methods per anatomical site. Only sites with at least 5% overall frequency in the included studies are displayed.

# 5. Discussion

We reviewed TL studies using CNN models well-trained on the ImageNet dataset for medical image analysis. We identified the most prevalent approaches regarding model selection, data augmentation, and visualization according to image modality and anatomical site. Previous reviews on medical imaging analysis covered the literature up to early 2017 [7] and late 2017 [8]. Those reviews included only a few TL studies using ImageNet. On the other hand, the majority (85%) of the studies included in the present review have been published after 2018. Therefore, we provide a critical update of the state-of-the-art in transfer learning methods for medical image analysis using ImageNet. Our findings can be used to help guide researchers to identify potential optimal approaches to specific medical image analysis problems as well as areas that warrant further research. These findings and research gaps are summarized in Table 4.

## 5.1. Transfer learning methods

From the imaging modality perspective, Inception models were the most frequently used for studies that analyzed X-Rays, endoscopic images (e.g., [62,64,65]), and ultrasound images (e.g., [55,57,58]), suggesting that wide networks (instead of deep networks) with inception modules benefiting from different kernel sizes may be more effective for these type of images. A few benchmarking studies comparing Inception models against very deep networks for these image types support this hypothesis (e.g., [27,43]). Most studies on skin lesion (43%)[89–91], fundus (42%) [77–81] and OCT images (50%) [92–94] showed that VGGNet obtained adequate performance, suggesting that shallow CNN models with multiple small kernel sizes may be optimal for processing these images. It is possible that small kernel sizes help capture detailed changes in images more accurately. Although a few studies have shown the better performance of shallow networks of VGGNet over deeper CNN models (e.g., [90,94]), and small kernel size over large kernel size (e.g., [78,80]), further research is needed with other deeper CNN models to confirm this hypothesis. GoogLeNet and AlexNet were the most prevalent approaches among studies that analyzed MRIs, suggesting that adequate accuracy can be achieved for these types of images without relying on very deep CNN models.

**Table 4**: The most prevalent methodological approaches and research gaps in the literature for various imaging modalities and anatomical sites. For the findings that were common among different imaging modalities and anatomical sites, one merged cell is used with those findings listed.

| Imaging modality | Anatomical site | CNN model | Data size and data collection method | Transfer learning approach | Final classifier | Visualization approach |
|---|---|---|---|---|---|---|
| MRI | Breast | Most prevalent:<br>• Wide networks.<br>Research gaps:<br>• Benchmarking is needed to find optimal models for different tasks. | Most prevalent:<br>• Feature-extracting TL for smaller datasets, and fine-tuning TL for larger datasets.<br><br>• Large datasets have been achieved by either collecting more labeled data or using data augmentation.<br><br>Research gaps:<br>• Find optimal dataset size thresholds for each medical image analysis problem.<br><br>• Investigate data augmentation methods other than image modification (e.g. image rotation, translation), such as generative adversarial network (GAN). | Most prevalent:<br>• Feature extracting TL approach for studies with less than 1,000 images after data augmentation.<br><br>• Fine-tuning TL approach for studies with more than 1,000 images.<br><br>Research gaps:<br>• Identify whether larger dataset or better choice of CNN model is the most important factor to optimize accuracy, time and memory for each TL approach. | Most prevalent:<br>• Studies that used a fine-tuning TL approach used fully connected layers (as opposed to traditional classifiers) more often than studies that used a feature extracting TL approach.<br><br>Research gaps:<br>• Benchmark traditional classifiers against fully connected layers when feature extracting TL approach is used. | Most prevalent:<br>• Heat maps were the most common approach in the studies that implemented a visualization technique.<br><br>Research gaps:<br>• Apply visualization techniques to provide insights on the decision-making process of the CNN model. |
| | Brain | Most prevalent:<br>• Shallow networks with large kernel sizes.<br>Research gaps:<br>• One benchmarking study compared different network types (no wide network included) for brain disease detection; [99] deep networks outperformed other approaches.<br>• Deep networks should be explored for other brain-related prediction tasks as well as other MRI anatomical sites (e.g., breast).<br>• Strong benchmarking is required, especially evaluating wide networks. | | | | |
| X-Ray | Breast | Most prevalent:<br>• Shallow networks with large kernel sizes.<br>Research gaps:<br>• Benchmarking is needed to find optimal models for different tasks. | | | | |
| | Lung | Most prevalent:<br>• Deep networks. Two robust benchmarking studies compared different network types; deep networks outperformed other approaches [130,131].<br>Research gaps:<br>• Deep networks should be explored for other X-Ray anatomical sites. | | | | |
| | Tooth | Most prevalent:<br>• Shallow networks with small kernel sizes.<br>Research gaps:<br>Benchmarking is needed to find optimal models for different tasks. | | | | |
| | Skeletal system | Most prevalent:<br>• Wide networks.<br>Research gaps:<br>• One benchmarking study compared different network types for bone age assessment as a numeric prediction task; [136] deep networks outperformed other approaches.<br>• Strong benchmarking for different nominal prediction tasks, which includes the majority of studies.<br>• Deep networks should be explored on other X-Ray skeletal system prediction tasks as well as other X-Ray anatomical sites. | | | | |
| CT | Various | Most prevalent:<br>• No prevalent network was found.<br>Research gaps:<br>• Few studies analyzed CT scans of different organs; little can be concluded about optimal CNN models for any anatomical site.<br>• Comprehensive benchmarking is critical to understand optimal models for different tasks. | | | | |
| Ultrasound | Various | Most prevalent:<br>• Wide networks.<br>Research gaps:<br>• One benchmarking study with a small dataset compared some network types (no wide network included) for breast lesion detection [134]; a deep network outperformed other approaches.<br>• Few studies analyzed ultrasound images of different organs; little can be concluded about optimal CNN models for any anatomical site.<br>• Stronger benchmarking that includes both wide and deep network types on large datasets is required. | | | | |
| Endoscopy | Stomach | Most prevalent:<br>• Wide networks.<br>Research gaps:<br>• One benchmarking study compared some network types (no deep network included) for gastric cancer | | | | |
| | | diagnosis [62]; wide networks outperformed other approaches.<br>• Stronger benchmarking is required, especially for evaluating the performance of deep networks. | | | | |
| Skin lesion | Skin | Most prevalent:<br>• Shallow networks with small kernel sizes.<br>Research gaps:<br>• One benchmarking study compared some network types (no deep network included) for melanoma diagnosis [66]; wide networks outperformed other approaches.<br>• Stronger benchmarking is required, especially for evaluating the performance of deep networks. | | | | |
| Fundus | Eye | Most prevalent:<br>• Shallow networks with small kernel sizes.<br>Research gaps:<br>• One benchmarking study compared all network types for diabetic retinopathy identification on OCT [135]; deep networks outperformed other approaches.<br>• Stronger benchmarking is needed both on Fundus and OCT images. More specifically, since the performance of shallow networks were close to deep networks in [135], optimal resource usage should be considered. | | | | |
| OCT | | | | | | |

Considering both anatomical site and imaging modality, Inception models (especially Inception-V3) were the most prevalent for analyzing X-Rays of the skeletal system (e.g., hip, knee, wrist) [39–41], suggesting the effectiveness of Inception models for this area. Similarly, GoogLeNet models combined with SVM classifiers were the most prevalent in breast MRI studies [49,50]. The effectiveness of wide networks (e.g., Inception models) for these anatomical sites and imaging modalities is supported by a few benchmarking studies that compared them against very deep networks (e.g., [48]), but more investigation is required. Most studies on brain MRI images [119–121,123] as well as breast X-Ray [114–116,118] images obtained adequate performance with AlexNet, which may indicate that shallow CNN models with large kernel sizes are optimal for those problems. Similarly, higher prevalence of VGGNet in tooth X-ray studies [68,70,72,73] suggests that shallow CNN models with small kernel sizes may be adequate for this kind of analysis. However, we did not find any benchmarking study focused on the analysis of tooth X-rays; further research with other CNN models is needed to confirm optimal models for the analysis of brain MRI and tooth X-ray. Models based on DenseNet were the most frequently used for studies that analyzed lung X-rays [128,130,131], suggesting that deeper CNN models are optimal for this problem, which is supported by two strong benchmarking studies ([130,131]). Finally, considering that only a few studies analyzed CT scans of different organs (e.g., tooth [60], prostate [126], and brain [9]), little can be concluded about optimal CNN models for these areas. We speculate that the small number of studies focused on CT images of those anatomical sites might result from lower clinical priority compared with other anatomical sites.

From the TL approach perspective (i.e., feature extracting or fine-tuning), the majority of studies with less than 1,000 images after data augmentation used a feature extracting TL approach, while the majority of studies with more than 1,000 images applied a fine-tuning TL approach. This finding is congruent with previous research, which showed similar preference patterns [139]. However, only few studies (e.g., [70,91]) applied both feature extracting and fine-tuning TL approaches on the same task, and compared their performance. Therefore, it is not clear whether larger data size (e.g., using data augmentation) or better choice of CNN model is the most important factor in determining accuracy and time and memory requirements.

Finally, for the final classifier, studies that used a fine-tuning TL approach used fully connected layers (as opposed to traditional classifiers) more often than studies that used a feature extracting TL approach (93% versus 68%). This choice may have been influenced by previous findings showing that feature extracting TL studies used smaller datasets compared to fine-tuning TL studies, since training the fully connected layers usually needs larger datasets compared to training traditional classifiers [139].

## 5.2. Dataset size and data augmentation methods

Data augmentation was more prevalent among studies that employed fine-tuning TL (72%) versus feature-extracting TL (15%). Moreover, in studies with smaller datasets (i.e., less than 1,000 images) most of the feature extracting TL studies did not perform data augmentation (78%) (e.g., [70,120]), while majority of the fine-tuning TL studies performed that (77%) (e.g., [123,127]). On the other hand, among studies with large datasets (i.e., more than 10,000) none of the feature extracting TL studies performed data augmentation, while still over half of the fine-tuning TL studies performed that (55%) (e.g., [64,104]). Congruent with previous findings [140], this suggests that feature-extracting TL can be done with smaller datasets, but fine-tuning TL requires larger datasets, which can be achieved by either collecting a large dataset (i.e., more labeled data) or using data augmentation.

Very few studies have reported performance results for various data sizes, or with and without data augmentation (e.g., [121]). Therefore, it is not clear to what extent the size of the dataset used in many studies (e.g., [64,133]) was essential to achieve the reported performance. Finding optimal thresholds for dataset size for each approach and medical image analysis problem is an important research gap because large datasets may not always be available. Another research gap is that only image modification (e.g. image rotation, translation) has been used as a method to create new data. Other methods to create high-quality synthetic images, such as generative adversarial network (GAN) [141], warrant investigation.

## 5.3. Classifier performance and visualization

The majority (65%) of the reviewed studies did not benchmark their CNN model against any other model, and 13% benchmarked against only one model. Since the majority of the studies in our systematic review were published after 2018, we can safely assume that investigators had access to current state-of-the-art of CNN models for benchmarking. In addition, studies comparing the performance of multiple models did not discuss the potential technical reason(s) that explain their findings. For instance, for diagnosing thyroid nodules, [84] has shown that VGGNet outperformed CNN models like ResNet and Inception, which have been developed after VGGNet, but no methodological discussion is provided. Also, there were many problems areas (e.g., CT scans for liver, tooth and brain) that had just one study with one single CNN model. Although all studies achieved adequate performance, we believe that there was possibly room for further performance improvement and/or complexity reduction if a wider range of CNN models had been tried in each study. Therefore, a stronger focus on systematic benchmarking through standardized methods is critical to better understand optimal approaches for each specific medical task. Moreover, based on the findings of seven benchmarking studies [99,109,130,131,134–136] on different imaging modalities and anatomical sites, deep CNN models always outperformed other CNN models. On the other hand, deep networks were the most prevalent approach only for lung X-rays. Thus, despite promising results, deep CNN models are understudied and should be further investigated with a variety of image modalities and anatomical sites.

Only 33% of the reviewed studies addressed CNN model visualization, mostly through heat maps (67%). This is an important research gap that warrants attention. CNN model visualization can provide insights on its decision-making process, which is crucial for establishing trust in the medical community [31]. Meaningful integration of CNN models in healthcare practice is highly unlikely, unless medical practitioners can understand, to some extent, its decision-making process. CNN model visualization can also benefit researchers as a diagnostic tool to further improve CNN methods [142–144].

This study had limitations. First, many of the initially selected studies were excluded due to lack of enough information for the review. Standard reporting is critical to improve the reproducibility of research in this area. For example, studies should include a clear description of the TL approach (i.e., feature-extracting or fine-tuning), including the final dataset size after augmentation, and report the final performance results for all models. Second, there were many problems areas that had just one study with a single CNN model for which we were not able to make any conclusions. Further research is needed to identify optimal methods for those areas. Third, due to the paucity of comparable benchmarking studies, our methodological implications need to be considered with caution. Further research is needed using standardized and replicable benchmarking methods to enhance comparability among studies. Finally, this study was limited to the use of well-trained CNN models on ImageNet in medical TL for image classification. Future reviews should focus on studies applying well-trained CNN models from other domains (based on non-ImageNet datasets) to medical image classification as well as other medical image tasks such as image segmentation.

## 6. Conclusion

We systematically reviewed TL studies that employed well-trained CNN models on the non-medical ImageNet dataset for medical image analysis. Regardless of data size, data augmentation method, CNN model and transfer learning approach, studies have generally achieved reasonable performance in their target task. This suggests that transfer learning using ImageNet, as a non-medical dataset, might be an effective way to approach medical tasks. This study identified the most prevalent tracks of implementation in the literature for data preparation, methodology selection and output evaluation for various medical image analysis tasks. Most prevalent models included wide CNN models using the Inception modules for ultrasound, endoscopy and skeletal system X-rays; shallow CNN models with large kernel size using AlexNet for brain MRIs and breast X-rays; deep CNN models with DenseNet for lung X-rays; and shallow CNN models with small kernel size using VGGNet models for eye (including fundus and OCT images), skin and dental X-rays. Feature-extracting TL was most prevalent with smaller datasets, while fine-tuning TL required larger datasets, sometimes achieved through data augmentation. Finally, fully connected layers for the final classification were also more prevalent with larger datasets.

We identified several research gaps existing in the TL studies on medical image analysis. First, the majority of studies did not benchmark their CNN models against other models. Stronger focus on systematic benchmarking through standardized methods is critical to understand optimal models for each medical imaging task. Second, based on the findings of seven benchmarking studies, deep models for a variety of image modalities and anatomical sites should be further investigated in future studies. Third, only a few studies applied and compared both feature extracting and fine-tuning TL approaches on the same task. Further research is required to identify whether larger data size or better choice of CNN model is the most important factor to optimize accuracy, time and memory. Fourth, because large datasets may not always be available, finding optimal dataset size thresholds for each medical image analysis problem is an important research gap. Fifth, instead of image modification (e.g. image rotation, translation) exploring other data augmentation methods such as generative adversarial network (GAN) warrants investigation. Sixth, the majority of studies did not apply visualization techniques to provide insights on the decision-making process of the CNN model. Meaningful integration of CNN models in healthcare practice is highly unlikely, unless medical practitioners can understand, to some extent, the rationale behind an algorithm's conclusion. Finally, in-depth analysis of studies within each individual imaging modality/anatomical site is needed to provide deeper insights into optimal methods and opportunities in each specific task.

# References

[1] K. Fukushima, Neocognitron: A self-organizing neural network model for a mechanism of pattern recognition unaffected by shift in position, Biol. Cybern. 36 (1980) 193–202. doi:10.1007/BF00344251.

[2] A. Borjali, A.F. Chen, O.K. Muratoglu, M.A. Morid, K.M. Varadarajan, Detecting total hip replacement prosthesis design on plain radiographs using deep convolutional neural network, J. Orthop. Res. (2020) jor.24617. doi:10.1002/jor.24617.

[3] A. Krizhevsky, I. Sutskever, G.E. Hinton, ImageNet Classification with Deep Convolutional Neural Networks, in: Adv. Neural Inf. Process. Syst., 2012: pp. 1097–1105.

[4] K. Simonyan, A. Zisserman, Two-Stream Convolutional Networks for Action Recognition in Videos, in: Adv. Neural Inf. Process. Syst., 2014: pp. 568–576.

[5] K. He, X. Zhang, S. Ren, J. Sun, Deep residual learning for image recognition, in: Proc. IEEE Comput. Soc. Conf. Comput. Vis. Pattern Recognit., IEEE Computer Society, 2016: pp. 770–778. doi:10.1109/CVPR.2016.90.

[6] M. Bakator, D. Radosav, Deep Learning and Medical Diagnosis: A Review of Literature, Multimodal Technol. Interact. 47 (2018). www.mdpi.com/journal/mti.

[7] G. Litjens, T. Kooi, B.E. Bejnordi, A.A.A. Setio, F. Ciompi, M. Ghafoorian, J.A.W.M. van der Laak, B. van Ginneken, C.I. Sánchez, A survey on deep learning in medical image analysis, Med. Image Anal. 42 (2017) 60–88. doi:10.1016/j.media.2017.07.005.

[8] V. Cheplygina, M. de Bruijne, J.P.W. Pluim, Not-so-supervised: A survey of semi-supervised, multi-instance, and transfer learning in medical image analysis., Med. Image Anal. 54 (2019) 280–296. doi:10.1016/j.media.2019.03.009.

[9] A.M. Dawud, K. Yurtkan, H. Oztoprak, Application of deep learning in neuroradiology: Brain haemorrhage classification using transfer learning, Comput. Intell. Neurosci. 2019 (2019). doi:10.1155/2019/4629859.

[10] Nitish Srivastava, Geoffrey Hinton, Alex Krizhevsky, Ilya Sutskever, Ruslan Salakhutdinov, Dropout: a simple way to prevent neural networks from overfitting, J. Mach. Learn. Res. 15 (2014) 1929–1958.

[11] J. Schmidhuber, Deep learning in neural networks: An overview, Neural Networks. 61 (2015) 85–117. doi:10.1016/J.NEUNET.2014.09.003.

[12] O. Russakovsky, J. Deng, H. Su, J. Krause, S. Satheesh, S. Ma, Z. Huang, A. Karpathy, A. Khosla, M. Bernstein, A.C. Berg, L. Fei-Fei, ImageNet Large Scale Visual Recognition Challenge, Int. J. Comput. Vis. 115 (2015) 211–252. doi:10.1007/s11263-015-0816-y.

[13] L.-J. Li, K. Li, F.F. Li, J. Deng, W. Dong, R. Socher, L. Fei-Fei, ImageNet: a Large-Scale Hierarchical Image Database Characterization of natural fibers View project hybrid intrusion detction systems View project ImageNet: A Large-Scale Hierarchical Image Database, in: IEEE Conf. Comput. Vis. Pattern Recognit., 2009: pp. 248–255. doi:10.1109/CVPR.2009.5206848.

[14] A. Khan, A. Sohail, U. Zahoora, A.S. Qureshi, A Survey of the Recent Architectures of Deep Convolutional Neural Networks, (2019).

[15] P. Ramachandran, B. Zoph, Q. V. Le, Searching for Activation Functions, (2017). http://arxiv.org/abs/1710.05941.

[16] K. Simonyan, A. Zisserman, Very Deep Convolutional Networks for Large-Scale Image Recognition, ArXiv Prepr. ArXiv1409.1556. (2014).

[17] Y. Jia, E. Shelhamer, J. Donahue, S. Karayev, J. Long, R. Girshick, S. Guadarrama, T. Darrell, Caffe: Convolutional architecture for fast feature embedding, in: Proc. 2014 ACM Conf. Multimed., Association for Computing Machinery, Inc, 2014: pp. 675–678. doi:10.1145/2647868.2654889.

[18] M.D. Zeiler, R. Fergus, Visualizing and understanding convolutional networks, in: Eur. Conf. Comput. Vis., Springer Verlag, 2014: pp. 818–833. doi:10.1007/978-3-319-10590-1_53.

[19] M.D. Zeiler, D. Krishnan, G.W. Taylor, R. Fergus, Deconvolutional networks, in: Proc. IEEE Comput. Soc. Conf. Comput. Vis. Pattern Recognit., 2010: pp. 2528–2535. doi:10.1109/CVPR.2010.5539957.

[20] C. Szegedy, W. Liu, Y. Jia, P. Sermanet, S. Reed, D. Anguelov, D. Erhan, V. Vanhoucke, A. Rabinovich, Going deeper with convolutions, in: Proc. IEEE Comput. Soc. Conf. Comput. Vis. Pattern Recognit., IEEE Computer Society, 2015: pp. 1–9. doi:10.1109/CVPR.2015.7298594.

[21] S. Ioffe, C. Szegedy, Batch normalization: Accelerating deep network training by reducing internal covariate shift, in: 32nd Int. Conf. Mach. Learn. ICML 2015, International Machine Learning Society (IMLS), 2015: pp. 448–456.

[22] C. Szegedy, V. Vanhoucke, S. Ioffe, J. Shlens, Rethinking the Inception Architecture for Computer Vision, in: IEEE Conf. Comput. Vis. Pattern Recognit., 2016: pp. 2818–2826.

[23]  C. Szegedy, S. Ioffe, V. Vanhoucke, A.A. Alemi, Inception-v4, Inception-ResNet and the Impact of Residual Connections on Learning, in: Thirty-First AAAI Conf. Artif. Intell., 2017. www.aaai.org.

[24]  A. Kensert, P.J. Harrison, O. Spjuth, Transfer Learning with Deep Convolutional Neural Networks for Classifying Cellular Morphological Changes, SLAS Discov. 24 (2019) 466–475. doi:10.1177/2472555218818756.

[25]  F. Chollet, Xception: Deep Learning with Depthwise Separable Convolutions, in: IEEE Conf. Comput. Vis. Pattern Recognit., 2017: pp. 1251–1258.

[26]  G. Huang, Z. Liu, L. van der Maaten, K.Q. Weinberger, Densely Connected Convolutional Networks, in: IEEE Conf. Comput. Vis. Pattern Recognit., 2017: pp. 4700–4708.

[27]  S. Zhou, X. Zhang, R. Zhang, Identifying Cardiomegaly in ChestX-ray8 Using Transfer Learning, in: Stud. Health Technol. Inform., IOS Press, 2019: pp. 482–486. doi:10.3233/SHTI190268.

[28]  S. Sharma, R. Mehra, Conventional Machine Learning and Deep Learning Approach for Multi-Classification of Breast Cancer Histopathology Images—a Comparative Insight, J. Digit. Imaging. (2020). doi:10.1007/s10278-019-00307-y.

[29]  J. Salamon, J.P. Bello, Deep Convolutional Neural Networks and Data Augmentation for Environmental Sound Classification, IEEE Signal Process. Lett. 24 (2017) 279–283. doi:10.1109/LSP.2017.2657381.

[30]  C. Shorten, T.M. Khoshgoftaar, A survey on Image Data Augmentation for Deep Learning, J. Big Data. 6 (2019). doi:10.1186/s40537-019-0197-0.

[31]  A. Borjali, A.F. Chen, O.K. Muratoglu, M.A. Morid, K.M. Varadarajan, Deep Learning in Orthopedics: How Do We Build Trust in the Machine?, Healthc. Transform. (2020) heat.2019.0006. doi:10.1089/heat.2019.0006.

[32]  Z. Qin, F. Yu, C. Liu, X. Chen, How convolutional neural networks see the world-a survey of convolutional neural network visualization methods, Inst. Math. Sci. 1 (2018) 149–180. doi:10.3934/mfc.2018008.

[33]  D. Silver, A. Huang, C.J. Maddison, A. Guez, L. Sifre, G. Van Den Driessche, J. Schrittwieser, I. Antonoglou, V. Panneershelvam, M. Lanctot, S. Dieleman, D. Grewe, J. Nham, N. Kalchbrenner, I. Sutskever, T. Lillicrap, M. Leach, K. Kavukcuoglu, T. Graepel, D. Hassabis, Mastering the game of Go with deep neural networks and tree search, Nature. 529 (2016) 484–489. doi:10.1038/nature16961.

[34]  M.D. Zeiler, G.W. Taylor, R. Fergus, Adaptive deconvolutional networks for mid and high level feature learning, in: Proc. IEEE Int. Conf. Comput. Vis., 2011: pp. 2018–2025. doi:10.1109/ICCV.2011.6126474.

[35]  B. Zhou, A. Khosla, A. Lapedriza, A. Oliva, A. Torralba, Learning Deep Features for Discriminative Localization, in: IEEE Conf. Comput. Vis. Pattern Recognit., 2016: pp. 2921–2929.

[36]  R.R. Selvaraju, M. Cogswell, A. Das, R. Vedantam, D. Parikh, D. Batra, Grad-CAM: Visual Explanations from Deep Networks via Gradient-based Localization, in: IEEE Int. Conf. Comput. Vis., 2017: pp. 618–626.

[37]  A.C. Tricco, E. Lillie, W. Zarin, K.K. O'Brien, H. Colquhoun, D. Levac, D. Moher, M.D.J. Peters, T. Horsley, L. Weeks, S. Hempel, E.A. Akl, C. Chang, J. McGowan, L. Stewart, L. Hartling, A. Aldcroft, M.G. Wilson, C. Garritty, S. Lewin, C.M. Godfrey, M.T. MacDonald, E. V. Langlois, K. Soares-Weiser, J. Moriarty, T. Clifford, Ö. Tunçalp, S.E. Straus, PRISMA extension for scoping reviews (PRISMA-ScR): Checklist and explanation, Ann. Intern. Med. 169 (2018) 467–473. doi:10.7326/M18-0850.

[38]  H. Arksey, L. O'Malley, Scoping studies: Towards a methodological framework, Int. J. Soc. Res. Methodol. Theory Pract. 8 (2005) 19–32. doi:10.1080/1364557032000119616.

[39]  A.Z. Abidin, B. Deng, A.M. DSouza, M.B. Nagarajan, P. Coan, A. Wismüller, Deep transfer learning for characterizing chondrocyte patterns in phase contrast X-Ray computed tomography images of the human patellar cartilage, Comput. Biol. Med. 95 (2018) 24–33. doi:10.1016/j.compbiomed.2018.01.008.

[40]  D.H. Kim, T. MacKinnon, Artificial intelligence in fracture detection: transfer learning from deep convolutional neural networks, Clin. Radiol. 73 (2018) 439–445. doi:10.1016/j.crad.2017.11.015.

[41]  J.S. Yu, S.M. Yu, B.S. Erdal, M. Demirer, V. Gupta, M. Bigelow, A. Salvador, T. Rink, S.S. Lenobel, L.M. Prevedello, R.D. White, Detection and localisation of hip fractures on anteroposterior radiographs with artificial intelligence: proof of concept, Clin. Radiol. (2019). doi:10.1016/j.crad.2019.10.022.

[42]  J.H. Lee, D.H. Kim, S.N. Jeong, S.H. Choi, Detection and diagnosis of dental caries using a deep learning-based convolutional neural network algorithm, J. Dent. 77 (2018) 106–111. doi:10.1016/j.jdent.2018.07.015.

[43]  F. Jiang, H. Liu, S. Yu, Y. Xie, Breast mass lesion classification in mammograms by transfer learning, in: ACM Int. Conf. Proceeding Ser., Association for Computing Machinery, 2017: pp. 59–62. doi:10.1145/3035012.3035022.

[44]  Y. Mednikov, S. Nehemia, B. Zheng, O. Benzaquen, D. Lederman, Transfer Representation Learning using Inception-V3 for the Detection of Masses in Mammography, in: Proc. Annu. Int. Conf. IEEE Eng.

Med. Biol. Soc. EMBS, Institute of Electrical and Electronics Engineers Inc., 2018: pp. 2587–2590. doi:10.1109/EMBC.2018.8512750.

[45] D. Arefan, A.A. Mohamed, W.A. Berg, M.L. Zuley, J.H. Sumkin, S. Wu, Deep learning modeling using normal mammograms for predicting breast cancer risk, Med. Phys. 47 (2020) 110–118. doi:10.1002/mp.13886.

[46] A. Rajkomar, S. Lingam, A.G. Taylor, M. Blum, J. Mongan, High-Throughput Classification of Radiographs Using Deep Convolutional Neural Networks, J. Digit. Imaging. 30 (2017) 95–101. doi:10.1007/s10278-016-9914-9.

[47] S. Deepak, P.M. Ameer, Brain tumor classification using deep CNN features via transfer learning, Comput. Biol. Med. 111 (2019). doi:10.1016/j.compbiomed.2019.103345.

[48] A.L. Dallora, J.S. Berglund, M. Brogren, O. Kvist, S. Diaz Ruiz, A. Dübbel, P. Anderberg, Age Assessment of Youth and Young Adults Using Magnetic Resonance Imaging of the Knee: A Deep Learning Approach., JMIR Med. Informatics. 7 (2019) e16291. doi:10.2196/16291.

[49] Z. Zhu, E. Albadawy, A. Saha, J. Zhang, M.R. Harowicz, M.A. Mazurowski, Deep learning for identifying radiogenomic associations in breast cancer, Comput. Biol. Med. 109 (2019) 85–90. doi:10.1016/j.compbiomed.2019.04.018.

[50] Z. Zhu, M. Harowicz, J. Zhang, A. Saha, L.J. Grimm, E.S. Hwang, M.A. Mazurowski, Deep learning analysis of breast MRIs for prediction of occult invasive disease in ductal carcinoma in situ, Comput. Biol. Med. 115 (2019). doi:10.1016/j.compbiomed.2019.103498.

[51] Y. Yang, L.F. Yan, X. Zhang, Y. Han, H.Y. Nan, Y.C. Hu, B. Hu, S.L. Yan, J. Zhang, D.L. Cheng, X.W. Ge, G. Bin Cui, D. Zhao, W. Wang, Glioma grading on conventional MR images: A deep learning study with transfer learning, Front. Neurosci. 12 (2018). doi:10.3389/fnins.2018.00804.

[52] F. Li, Z. Liu, H. Chen, M. Jiang, X. Zhang, Z. Wu, Automatic detection of diabetic retinopathy in retinal fundus photographs based on deep learning algorithm, Transl. Vis. Sci. Technol. 8 (2019). doi:10.1167/tvst.8.6.4.

[53] F. Arcadu, F. Benmansour, A. Maunz, J. Michon, Z. Haskova, D. McClintock, A.P. Adamis, J.R. Willis, M. Prunotto, Deep learning predicts OCT measures of diabetic macular thickening from color fundus photographs, Investig. Ophthalmol. Vis. Sci. 60 (2019) 852–857. doi:10.1167/iovs.18-25634.

[54] T. Xiao, L. Liu, K. Li, W. Qin, S. Yu, Z. Li, Comparison of Transferred Deep Neural Networks in Ultrasonic Breast Masses Discrimination, Biomed Res. Int. 2018 (2018). doi:10.1155/2018/4605191.

[55] J. Song, Y.J. Chai, H. Masuoka, S.W. Park, S.J. Kim, J.Y. Choi, H.J. Kong, K.E. Lee, J. Lee, N. Kwak, K.H. Yi, A. Miyauchi, Ultrasound image analysis using deep learning algorithm for the diagnosis of thyroid nodules, Medicine (Baltimore). 98 (2019) e15133. doi:10.1097/MD.0000000000015133.

[56] L.-Y. Xue, Z.-Y. Jiang, T.-T. Fu, Q.-M. Wang, Y.-L. Zhu, M. Dai, W.-P. Wang, J.-H. Yu, H. Ding, Transfer learning radiomics based on multimodal ultrasound imaging for staging liver fibrosis, Eur. Radiol. (2020). doi:10.1007/s00330-019-06595-w.

[57] J. Chi, E. Walia, P. Babyn, J. Wang, G. Groot, M. Eramian, Thyroid Nodule Classification in Ultrasound Images by Fine-Tuning Deep Convolutional Neural Network, J. Digit. Imaging. 30 (2017) 477–486. doi:10.1007/s10278-017-9997-y.

[58] Q. Guan, Y. Wang, J. Du, Y. Qin, H. Lu, J. Xiang, F. Wang, Deep learning based classification of ultrasound images for thyroid nodules: a large scale of pilot study, Ann. Transl. Med. 7 (2019) 137–137. doi:10.21037/atm.2019.04.34.

[59] T. Fujioka, K. Kubota, M. Mori, Y. Kikuchi, L. Katsuta, M. Kasahara, G. Oda, T. Ishiba, T. Nakagawa, U. Tateishi, Distinction between benign and malignant breast masses at breast ultrasound using deep learning method with convolutional neural network, Jpn. J. Radiol. 37 (2019) 466–472. doi:10.1007/s11604-019-00831-5.

[60] J.H. Lee, D.H. Kim, S.N. Jeong, Diagnosis of cystic lesions using panoramic and cone beam computed tomographic images based on deep learning neural network, Oral Dis. (2019). doi:10.1111/odi.13223.

[61] N.I. Chowdhury, T.L. Smith, R.K. Chandra, J.H. Turner, Automated classification of osteomeatal complex inflammation on computed tomography using convolutional neural networks, Int. Forum Allergy Rhinol. 9 (2019) 46–52. doi:10.1002/alr.22196.

[62] X. Liu, C. Wang, Y. Hu, Z. Zeng, J. Bai, G. Liao, Transfer Learning with Convolutional Neural Network for Early Gastric Cancer Classification on Magnifiying Narrow-Band Imaging Images, in: Proc. - Int. Conf. Image Process. ICIP, IEEE Computer Society, 2018: pp. 1388–1392. doi:10.1109/ICIP.2018.8451067.

[63] X. Li, H. Zhang, X. Zhang, H. Liu, G. Xie, Exploring transfer learning for gastrointestinal bleeding detection on small-size imbalanced endoscopy images, in: Proc. Annu. Int. Conf. IEEE Eng. Med. Biol. Soc. EMBS, Institute of Electrical and Electronics Engineers Inc., 2017: pp. 1994–1997.

[64] Y. Sakai, S. Takemoto, K. Hori, M. Nishimura, H. Ikematsu, T. Yano, H. Yokota, Automatic detection of early gastric cancer in endoscopic images using a transferring convolutional neural network, in: Proc. Annu. Int. Conf. IEEE Eng. Med. Biol. Soc. EMBS, Institute of Electrical and Electronics Engineers Inc., 2018: pp. 4138–4141. doi:10.1109/EMBC.2018.8513274.

[65] L. Li, Y. Chen, Z. Shen, X. Zhang, J. Sang, Y. Ding, X. Yang, J. Li, M. Chen, C. Jin, C. Chen, C. Yu, Convolutional neural network for the diagnosis of early gastric cancer based on magnifying narrow band imaging, Gastric Cancer. 23 (2020) 126–132. doi:10.1007/s10120-019-00992-2.

[66] X. Cui, R. Wei, L. Gong, R. Qi, Z. Zhao, H. Chen, K. Song, A.A.A. Abdulrahman, Y. Wang, J.Z.S. Chen, S. Chen, Y. Zhao, X. Gao, Assessing the effectiveness of artificial intelligence methods for melanoma: A retrospective review, J. Am. Acad. Dermatol. 81 (2019) 1176–1180. doi:10.1016/j.jaad.2019.06.042.

[67] D.S. Kermany, M. Goldbaum, W. Cai, C.C.S. Valentim, H. Liang, S.L. Baxter, A. McKeown, G. Yang, X. Wu, F. Yan, J. Dong, M.K. Prasadha, J. Pei, M.Y.L. Ting, J. Zhu, C. Li, S. Hewett, J. Dong, I. Ziyar, A. Shi, R. Zhang, L. Zheng, R. Hou, W. Shi, X. Fu, Y. Duan, V.A.N. Huu, C. Wen, E.D. Zhang, C.L. Zhang, O. Li, X. Wang, M.A. Singer, X. Sun, J. Xu, A. Tafreshi, M.A. Lewis, H. Xia, K. Zhang, Identifying Medical Diagnoses and Treatable Diseases by Image-Based Deep Learning, Cell. 172 (2018) 1122-1131.e9. doi:10.1016/J.CELL.2018.02.010.

[68] W. Poedjiastoeti, S. Suebnukarn, Application of convolutional neural network in the diagnosis of Jaw tumors, Healthc. Inform. Res. 24 (2018) 236–241. doi:10.4258/hir.2018.24.3.236.

[69] M. Ahsan, R. Gomes, A. Denton, Application of a convolutional neural network using transfer learning for tuberculosis detection, in: IEEE Int. Conf. Electro Inf. Technol., IEEE Computer Society, 2019: pp. 427–433. doi:10.1109/EIT.2019.8833768.

[70] S.A. Prajapati, R. Nagaraj, S. Mitra, Classification of dental diseases using CNN and transfer learning, in: 5th Int. Symp. Comput. Bus. Intell. ISCBI 2017, Institute of Electrical and Electronics Engineers Inc., 2017: pp. 70–74. doi:10.1109/ISCBI.2017.8053547.

[71] S. Yune, H. Lee, M. Kim, S.H. Tajmir, M.S. Gee, S. Do, Beyond Human Perception: Sexual Dimorphism in Hand and Wrist Radiographs Is Discernible by a Deep Learning Model., J. Digit. Imaging. 32 (2019) 665–671. doi:10.1007/s10278-018-0148-x.

[72] Lee, Jung, Ryu, Shin, Choi, Evaluation of Transfer Learning with Deep Convolutional Neural Networks for Screening Osteoporosis in Dental Panoramic Radiographs, J. Clin. Med. 9 (2020) 392. doi:10.3390/jcm9020392.

[73] J.H. Lee, D.H. Kim, S.N. Jeong, S.H. Choi, Diagnosis and prediction of periodontally compromised teeth using a deep learning-based convolutional neural network algorithm, J. Periodontal Implant Sci. 48 (2018) 114–123. doi:10.5051/jpis.2018.48.2.114.

[74] Z.N.K. Swati, Q. Zhao, M. Kabir, F. Ali, Z. Ali, S. Ahmed, J. Lu, Content-Based Brain Tumor Retrieval for MR Images Using Transfer Learning, IEEE Access. 7 (2019) 17809–17822. doi:10.1109/ACCESS.2019.2892455.

[75] T. Langner, J. Wikstrom, T. Bjerner, H. Ahlstrom, J. Kullberg, Identifying morphological indicators of aging with neural networks on large-scale whole-body MRI, IEEE Trans. Med. Imaging. (2019) 1–1. doi:10.1109/tmi.2019.2950092.

[76] N.M. Khan, N. Abraham, M. Hon, Transfer Learning with Intelligent Training Data Selection for Prediction of Alzheimer's Disease, IEEE Access. 7 (2019) 72726–72735. doi:10.1109/ACCESS.2019.2920448.

[77] J.J. Gómez-Valverde, A. Antón, G. Fatti, B. Liefers, A. Herranz, A. Santos, C.I. Sánchez, M.J. Ledesma-Carbayo, Automatic glaucoma classification using color fundus images based on convolutional neural networks and transfer learning., Biomed. Opt. Express. 10 (2019) 892–913. doi:10.1364/BOE.10.000892.

[78] J.Y. Choi, T.K. Yoo, J.G. Seo, J. Kwak, T.T. Um, T.H. Rim, Multi-categorical deep learning neural network to classify retinal images: A pilot study employing small database, PLoS One. 12 (2017). doi:10.1371/journal.pone.0187336.

[79] X. Li, T. Pang, B. Xiong, W. Liu, P. Liang, T. Wang, Convolutional neural networks based transfer learning for diabetic retinopathy fundus image classification, in: Proc. - 2017 10th Int. Congr. Image Signal Process. Biomed. Eng. Informatics, CISP-BMEI 2017, Institute of Electrical and Electronics Engineers Inc., 2018: pp. 1–11. doi:10.1109/CISP-BMEI.2017.8301998.

[80] Y. Zhang, L. Wang, Z. Wu, J. Zeng, Y. Chen, R. Tian, J. Zhao, G. Zhang, Development of an Automated Screening System for Retinopathy of Prematurity Using a Deep Neural Network for Wide-Angle Retinal Images, IEEE Access. 7 (2019) 10232–10241. doi:10.1109/ACCESS.2018.2881042.

[81] D. Nagasato, H. Tabuchi, H. Ohsugi, H. Masumoto, H. Enno, N. Ishitobi, T. Sonobe, M. Kameoka, M. Niki, Y. Mitamura, Deep-learning classifier with ultrawide-field fundus ophthalmoscopy for detecting

branch retinal vein occlusion., Int. J. Ophthalmol. 12 (2019) 94–99. doi:10.18240/ijo.2019.01.15.

[82] M. Byra, M. Galperin, H. Ojeda-Fournier, L. Olson, M. O'Boyle, C. Comstock, M. Andre, Breast mass classification in sonography with transfer learning using a deep convolutional neural network and color conversion, Med. Phys. 46 (2019) 746–755. doi:10.1002/mp.13361.

[83] P.M. Cheng, H.S. Malhi, Transfer Learning with Convolutional Neural Networks for Classification of Abdominal Ultrasound Images, J. Digit. Imaging. 30 (2017) 234–243. doi:10.1007/s10278-016-9929-2.

[84] P. Qin, K. Wu, Y. Hu, J. Zeng, X. Chai, Diagnosis of benign and malignant thyroid nodules using combined conventional ultrasound and ultrasound elasticity imaging, IEEE J. Biomed. Heal. Informatics. (2019) 1–1. doi:10.1109/JBHI.2019.2950994.

[85] M. Nishio, O. Sugiyama, M. Yakami, S. Ueno, T. Kubo, T. Kuroda, K. Togashi, Computer-aided diagnosis of lung nodule classification between benign nodule, primary lung cancer, and metastatic lung cancer at different image size using deep convolutional neural network with transfer learning, PLoS One. 13 (2018). doi:10.1371/journal.pone.0200721.

[86] S. Belharbi, C. Chatelain, R. Hérault, S. Adam, S. Thureau, M. Chastan, R. Modzelewski, Spotting L3 slice in CT scans using deep convolutional network and transfer learning, Comput. Biol. Med. 87 (2017) 95–103. doi:10.1016/j.compbiomed.2017.05.018.

[87] M. Santin, C. Brama, H. Théro, E. Ketheeswaran, I. El-Karoui, F. Bidault, R. Gillet, P. Gondim Teixeira, A. Blum, Detecting abnormal thyroid cartilages on CT using deep learning, Diagn. Interv. Imaging. 100 (2019) 251–257. doi:10.1016/j.diii.2019.01.008.

[88] G. Wimmer, A. Vécsei, A. Uhl, CNN transfer learning for the automated diagnosis of celiac disease, in: 2016 6th Int. Conf. Image Process. Theory, Tools Appl. IPTA 2016, Institute of Electrical and Electronics Engineers Inc., 2017. doi:10.1109/IPTA.2016.7821020.

[89] C. Yu, S. Yang, W. Kim, J. Jung, K.Y. Chung, S.W. Lee, B. Oh, Acral melanoma detection using a convolutional neural network for dermoscopy images, PLoS One. 13 (2018) e0193321. doi:10.1371/journal.pone.0193321.

[90] A. Kwasigroch, A. Mikołajczyk, M. Grochowski, Deep neural networks approach to skin lesions classification - A comparative analysis, in: 2017 22nd Int. Conf. Methods Model. Autom. Robot. MMAR 2017, Institute of Electrical and Electronics Engineers Inc., 2017: pp. 1069–1074. doi:10.1109/MMAR.2017.8046978.

[91] A. Romero Lopez, X. Giro-I-Nieto, J. Burdick, O. Marques, Skin lesion classification from dermoscopic images using deep learning techniques, in: Proc. 13th IASTED Int. Conf. Biomed. Eng. BioMed 2017, Institute of Electrical and Electronics Engineers Inc., 2017: pp. 49–54. doi:10.2316/P.2017.852-053.

[92] G. An, K. Omodaka, K. Hashimoto, S. Tsuda, Y. Shiga, N. Takada, T. Kikawa, H. Yokota, M. Akiba, T. Nakazawa, Glaucoma Diagnosis with Machine Learning Based on Optical Coherence Tomography and Color Fundus Images, J. Healthc. Eng. 2019 (2019). doi:10.1155/2019/4061313.

[93] F. Li, H. Chen, Z. Liu, X. Zhang, Z. Wu, Fully automated detection of retinal disorders by image-based deep learning, Graefe's Arch. Clin. Exp. Ophthalmol. 257 (2019) 495–505. doi:10.1007/s00417-018-04224-8.

[94] K.A. Thakoor, X. Li, E. Tsamis, P. Sajda, D.C. Hood, Enhancing the Accuracy of Glaucoma Detection from OCT Probability Maps using Convolutional Neural Networks, in: Proc. Annu. Int. Conf. IEEE Eng. Med. Biol. Soc. EMBS, Institute of Electrical and Electronics Engineers Inc., 2019: pp. 2036–2040. doi:10.1109/EMBC.2019.8856899.

[95] P.H. Yi, A. Lin, J. Wei, A.C. Yu, H.I. Sair, F.K. Hui, G.D. Hager, S.C. Harvey, Deep-Learning-Based Semantic Labeling for 2D Mammography and Comparison of Complexity for Machine Learning Tasks, J. Digit. Imaging. 32 (2019) 565–570. doi:10.1007/s10278-019-00244-w.

[96] P.H. Yi, J. Wei, T.K. Kim, H.I. Sair, F.K. Hui, G.D. Hager, J. Fritz, J.K. Oni, Automated detection & classification of knee arthroplasty using deep learning, Knee. (2019). doi:10.1016/j.knee.2019.11.020.

[97] P.H. Yi, T.K. Kim, J. Wei, J. Shin, F.K. Hui, H.I. Sair, G.D. Hager, J. Fritz, Automated semantic labeling of pediatric musculoskeletal radiographs using deep learning, Pediatr. Radiol. 49 (2019) 1066–1070. doi:10.1007/s00247-019-04408-2.

[98] P. Korfiatis, T.L. Kline, D.H. Lachance, I.F. Parney, J.C. Buckner, B.J. Erickson, Residual Deep Convolutional Neural Network Predicts MGMT Methylation Status, J. Digit. Imaging. 30 (2017) 622–628. doi:10.1007/s10278-017-0009-z.

[99] M. Talo, O. Yildirim, U.B. Baloglu, G. Aydin, U.R. Acharya, Convolutional neural networks for multi-class brain disease detection using MRI images, Comput. Med. Imaging Graph. 78 (2019). doi:10.1016/j.compmedimag.2019.101673.

[100] Y. Gao, Y. Zhang, H. Wang, X. Guo, J. Zhang, Decoding Behavior Tasks from Brain Activity Using Deep Transfer Learning, IEEE Access. 7 (2019) 43222–43232. doi:10.1109/ACCESS.2019.2907040.

[101] X. Zhong, R. Cao, S. Shakeri, F. Scalzo, Y. Lee, D.R. Enzmann, H.H. Wu, S.S. Raman, K. Sung, Deep transfer learning-based prostate cancer classification using 3 Tesla multi-parametric MRI, Abdom. Radiol. 44 (2019) 2030–2039. doi:10.1007/s00261-018-1824-5.

[102] T.Y.A. Liu, D.S.W. Ting, P.H. Yi, J. Wei, H. Zhu, P.S. Subramanian, T. Li, F.K. Hui, G.D. Hager, N.R. Miller, Deep Learning and Transfer Learning for Optic Disc Laterality Detection: : Implications for Machine Learning in Neuro-Ophthalmology, J. Neuro-Ophthalmology. (2019) 1. doi:10.1097/WNO.0000000000000827.

[103] R. Hemelings, B. Elen, J. Barbosa-Breda, S. Lemmens, M. Meire, S. Pourjavan, E. Vandewalle, S. Van de Veire, M.B. Blaschko, P. De Boever, I. Stalmans, Accurate prediction of glaucoma from colour fundus images with a convolutional neural network that relies on active and transfer learning, Acta Ophthalmol. (2019). doi:10.1111/aos.14193.

[104] M. Christopher, A. Belghith, C. Bowd, J.A. Proudfoot, M.H. Goldbaum, R.N. Weinreb, C.A. Girkin, J.M. Liebmann, L.M. Zangwill, Performance of Deep Learning Architectures and Transfer Learning for Detecting Glaucomatous Optic Neuropathy in Fundus Photographs, Sci. Rep. 8 (2018). doi:10.1038/s41598-018-35044-9.

[105] J. Li, X. Xu, Y. Guan, A. Imran, B. Liu, L. Zhang, J.J. Yang, Q. Wang, L. Xie, Automatic Cataract Diagnosis by Image-Based Interpretability, in: Proc. - 2018 IEEE Int. Conf. Syst. Man, Cybern. SMC 2018, Institute of Electrical and Electronics Engineers Inc., 2019: pp. 3964–3969. doi:10.1109/SMC.2018.00672.

[106] C.-C. Kuo, C.-M. Chang, K.-T. Liu, W.-K. Lin, H.-Y. Chiang, C.-W. Chung, M.-R. Ho, P.-R. Sun, R.-L. Yang, K.-T. Chen, Automation of the kidney function prediction and classification through ultrasound-based kidney imaging using deep learning, Npj Digit. Med. 2 (2019). doi:10.1038/s41746-019-0104-2.

[107] J.H. Lee, E.J. Ha, J.H. Kim, Application of deep learning to the diagnosis of cervical lymph node metastasis from thyroid cancer with CT, Eur. Radiol. 29 (2019) 5452–5457. doi:10.1007/s00330-019-06098-8.

[108] J. Peng, S. Kang, Z. Ning, H. Deng, J. Shen, Y. Xu, J. Zhang, W. Zhao, X. Li, W. Gong, J. Huang, L. Liu, Residual convolutional neural network for predicting response of transarterial chemoembolization in hepatocellular carcinoma from CT imaging, Eur. Radiol. (2019). doi:10.1007/s00330-019-06318-1.

[109] R.V.M. Da Nóbrega, S.A. Peixoto, S.P.P. Da Silva, P.P.R. Filho, Lung Nodule Classification via Deep Transfer Learning in CT Lung Images, in: Proc. - IEEE Symp. Comput. Med. Syst., Institute of Electrical and Electronics Engineers Inc., 2018: pp. 244–249. doi:10.1109/CBMS.2018.00050.

[110] J.H. Lee, Y.J. Kim, Y.W. Kim, S. Park, Y. i. Choi, Y.J. Kim, D.K. Park, K.G. Kim, J.W. Chung, Spotting malignancies from gastric endoscopic images using deep learning, Surg. Endosc. 33 (2019) 3790–3797. doi:10.1007/s00464-019-06677-2.

[111] Y. Zhu, Q.C. Wang, M.D. Xu, Z. Zhang, J. Cheng, Y.S. Zhong, Y.Q. Zhang, W.F. Chen, L.Q. Yao, P.H. Zhou, Q.L. Li, Application of convolutional neural network in the diagnosis of the invasion depth of gastric cancer based on conventional endoscopy, Gastrointest. Endosc. 89 (2019) 806-815.e1. doi:10.1016/j.gie.2018.11.011.

[112] S. Hosseinzadeh Kassani, P. Hosseinzadeh Kassani, A comparative study of deep learning architectures on melanoma detection, Tissue Cell. 58 (2019) 76–83. doi:10.1016/j.tice.2019.04.009.

[113] W. Lu, Y. Tong, Y. Yu, Y. Xing, C. Chen, Y. Shen, Deep learning-based automated classification of multi-categorical abnormalities from optical coherence tomography images, Transl. Vis. Sci. Technol. 7 (2018). doi:10.1167/tvst.7.6.41.

[114] B.Q. Huynh, H. Li, M.L. Giger, Digital mammographic tumor classification using transfer learning from deep convolutional neural networks., J. Med. Imaging (Bellingham, Wash.). 3 (2016) 034501. doi:10.1117/1.JMI.3.3.034501.

[115] X. Zhang, Y. Zhang, E.Y. Han, N. Jacobs, Q. Han, X. Wang, J. Liu, Classification of whole mammogram and tomosynthesis images using deep convolutional neural networks, IEEE Trans. Nanobioscience. 17 (2018) 237–242. doi:10.1109/TNB.2018.2845103.

[116] H. Li, M.L. Giger, B.Q. Huynh, N.O. Antropova, Deep learning in breast cancer risk assessment: evaluation of convolutional neural networks on a clinical dataset of full-field digital mammograms., J. Med. Imaging (Bellingham, Wash.). 4 (2017) 041304. doi:10.1117/1.JMI.4.4.041304.

[117] A. Abbas, M.M. Abdelsamea, Learning Transformations for Automated Classification of Manifestation of Tuberculosis using Convolutional Neural Network, in: Proc. - 2018 13th Int. Conf. Comput. Eng. Syst. ICCES 2018, Institute of Electrical and Electronics Engineers Inc., 2019: pp. 122–126. doi:10.1109/ICCES.2018.8639200.

[118] D.A. Ragab, M. Sharkas, S. Marshall, J. Ren, Breast cancer detection using deep convolutional neural networks and support vector machines, PeerJ. 2019 (2019) e6201. doi:10.7717/peerj.6201.

[119] C. Zhang, K. Qiao, L. Wang, L. Tong, G. Hu, R.Y. Zhang, B. Yan, A visual encoding model based on deep neural networks and transfer learning for brain activity measured by functional magnetic resonance imaging, J. Neurosci. Methods. 325 (2019). doi:10.1016/j.jneumeth.2019.108318.

[120] M. Maqsood, F. Nazir, U. Khan, F. Aadil, H. Jamal, I. Mehmood, O.Y. Song, Transfer learning assisted classification and detection of alzheimer's disease stages using 3D MRI scans, Sensors (Switzerland). 19 (2019). doi:10.3390/s19112645.

[121] S.H. Wang, S. Xie, X. Chen, D.S. Guttery, C. Tang, J. Sun, Y.D. Zhang, Alcoholism identification based on an Alexnet transfer learning model, Front. Psychiatry. 10 (2019). doi:10.3389/fpsyt.2019.00205.

[122] Y. Yuan, W. Qin, M. Buyyounouski, B. Ibragimov, S. Hancock, B. Han, L. Xing, Prostate cancer classification with multiparametric MRI transfer learning model, Med. Phys. 46 (2019) 756–765. doi:10.1002/mp.13367.

[123] S. Afzal, M. Maqsood, F. Nazir, U. Khan, F. Aadil, K.M. Awan, I. Mehmood, O.-Y. Song, A Data Augmentation-Based Framework to Handle Class Imbalance Problem for Alzheimer's Stage Detection, IEEE Access. 7 (2019) 115528–115539. doi:10.1109/access.2019.2932786.

[124] A. Li, J. Cheng, D.W.K. Wong, J. Liu, Integrating holistic and local deep features for glaucoma classification, in: Proc. Annu. Int. Conf. IEEE Eng. Med. Biol. Soc. EMBS, Institute of Electrical and Electronics Engineers Inc., 2016: pp. 1328–1331. doi:10.1109/EMBC.2016.7590952.

[125] H.C. Shin, H.R. Roth, M. Gao, L. Lu, Z. Xu, I. Nogues, J. Yao, D. Mollura, R.M. Summers, Deep Convolutional Neural Networks for Computer-Aided Detection: CNN Architectures, Dataset Characteristics and Transfer Learning, IEEE Trans. Med. Imaging. 35 (2016) 1285–1298. doi:10.1109/TMI.2016.2528162.

[126] T. Kajikawa, N. Kadoya, K. Ito, Y. Takayama, T. Chiba, S. Tomori, K. Takeda, K. Jingu, Automated prediction of dosimetric eligibility of patients with prostate cancer undergoing intensity-modulated radiation therapy using a convolutional neural network, Radiol. Phys. Technol. 11 (2018) 320–327. doi:10.1007/s12194-018-0472-3.

[127] K.M. Hosny, M.A. Kassem, M.M. Foaud, Classification of skin lesions using transfer learning and augmentation with Alex-net, PLoS One. 14 (2019). doi:10.1371/journal.pone.0217293.

[128] O. Gozes, H. Greenspan, Deep Feature Learning from a Hospital-Scale Chest X-ray Dataset with Application to TB Detection on a Small-Scale Dataset., Conf. Proc. ... Annu. Int. Conf. IEEE Eng. Med. Biol. Soc. IEEE Eng. Med. Biol. Soc. Annu. Conf. 2019 (2019) 4076–4079. doi:10.1109/EMBC.2019.8856729.

[129] H.J. Yu, S.R. Cho, M.J. Kim, W.H. Kim, J.W. Kim, J. Choi, Automated Skeletal Classification with Lateral Cephalometry Based on Artificial Intelligence, J. Dent. Res. (2020) 002203452090171. doi:10.1177/0022034520901715.

[130] Q.H. Nguyen, B.P. Nguyen, S.D. Dao, B. Unnikrishnan, R. Dhingra, S.R. Ravichandran, S. Satpathy, P.N. Raja, M.C.H. Chua, Deep Learning Models for Tuberculosis Detection from Chest X-ray Images, in: 2019 26th Int. Conf. Telecommun. ICT 2019, Institute of Electrical and Electronics Engineers Inc., 2019: pp. 381–385. doi:10.1109/ICT.2019.8798798.

[131] D. Varshni, K. Thakral, L. Agarwal, R. Nijhawan, A. Mittal, Pneumonia Detection Using CNN based Feature Extraction, in: Proc. 2019 3rd IEEE Int. Conf. Electr. Comput. Commun. Technol. ICECCT 2019, Institute of Electrical and Electronics Engineers Inc., 2019. doi:10.1109/ICECCT.2019.8869364.

[132] I. Pan, S. Agarwal, D. Merck, Generalizable Inter-Institutional Classification of Abnormal Chest Radiographs Using Efficient Convolutional Neural Networks, J. Digit. Imaging. 32 (2019) 888–896. doi:10.1007/s10278-019-00180-9.

[133] J.A. Dunnmon, D. Yi, C.P. Langlotz, C. Ré, D.L. Rubin, M.P. Lungren, Assessment of Convolutional Neural Networks for Automated Classification of Chest Radiographs, Radiology. 290 (2019) 537–544. doi:10.1148/radiol.2018181422.

[134] Z. Cao, L. Duan, G. Yang, T. Yue, Q. Chen, An experimental study on breast lesion detection and classification from ultrasound images using deep learning architectures, BMC Med. Imaging. 19 (2019). doi:10.1186/s12880-019-0349-x.

[135] K.T. Islam, S. Wijewickrema, S. O'Leary, Identifying diabetic retinopathy from OCT images using deep transfer learning with artificial neural networks, in: Proc. - IEEE Symp. Comput. Med. Syst., Institute of Electrical and Electronics Engineers Inc., 2019: pp. 281–286. doi:10.1109/CBMS.2019.00066.

[136] J. Han, Y. Jia, C. Zhao, F. Gou, Automatic Bone Age Assessment Combined with Transfer Learning and Support Vector Regression, in: Proc. - 9th Int. Conf. Inf. Technol. Med. Educ. ITME 2018, Institute of Electrical and Electronics Engineers Inc., 2018: pp. 61–66. doi:10.1109/ITME.2018.00025.

[137] M. Byra, G. Styczynski, C. Szmigielski, P. Kalinowski, Ł. Michałowski, R. Paluszkiewicz, B. Ziarkiewicz-Wróblewska, K. Zieniewicz, P. Sobieraj, A. Nowicki, Transfer learning with deep

convolutional neural network for liver steatosis assessment in ultrasound images, Int. J. Comput. Assist. Radiol. Surg. 13 (2018) 1895–1903. doi:10.1007/s11548-018-1843-2.

[138] H. Binol, A. Plotner, J. Sopkovich, B. Kaffenberger, M.K.K. Niazi, M.N. Gurcan, Ros-NET: A deep convolutional neural network for automatic identification of rosacea lesions, Ski. Res. Technol. (2019) srt.12817. doi:10.1111/srt.12817.

[139] D. Soekhoe, P. van der Putten, A. Plaat, On the impact of data set size in transfer learning using deep neural networks, in: Lect. Notes Comput. Sci. (Including Subser. Lect. Notes Artif. Intell. Lect. Notes Bioinformatics), Springer Verlag, 2016: pp. 50–60. doi:10.1007/978-3-319-46349-0_5.

[140] J. Cho, K. Lee, E. Shin, G. Choy, S. Do, How much data is needed to train a medical image deep learning system to achieve necessary high accuracy?, (2015). http://arxiv.org/abs/1511.06348.

[141] I.J. Goodfellow, J. Pouget-Abadie, M. Mirza, B. Xu, D. Warde-Farley, S. Ozair, A. Courville, Y. Bengio, Generative Adversarial Nets, in: Adv. Neural Inf. Process. Syst., 2014: pp. 2672–2680.

[142] L. Gordon, T. Grantcharov, F. Rudzicz, Explainable Artificial Intelligence for Safe Intraoperative Decision Support, JAMA Surg. 154 (2019) 1064–1065. doi:10.1001/jamasurg.2019.2821.

[143] M. Amin Morid, O.R. Liu Sheng, K. Kawamoto, S. Abdelrahman, Learning Hidden Patterns from Patient Multivariate Time Series Data Using Convolutional Neural Networks: A Case Study of Healthcare Cost Prediction, J. Biomed. Inform. (2020) 103565. doi:10.1016/j.jbi.2020.103565.

[144] S. Nundy, T. Montgomery, R.M. Wachter, Promoting trust between patients and physicians in the era of artificial intelligence, JAMA - J. Am. Med. Assoc. 322 (2019) 497–498. doi:10.1001/jama.2018.20563.

**Table S1**: Search strategy.

| Database | Search query |
|---|---|
| **MEDLINE** | ("transfer learning"[All Fields] OR "deep learning"[All Fields] OR "convolutional neural network"[All Fields] OR "convolutional neural networks"[All Fields]) AND ( "MRI"[All Fields] OR "MRIs"[All Fields] OR "Magnetic resonance images"[All Fields] OR "Magnetic resonance image"[All Fields] OR "MR image"[All Fields] OR "MR images"[All Fields] OR "CT"[All Fields] OR "CTs"[All Fields] OR "computed tomographic image"[All Fields] OR "computed tomographic images"[All Fields]  OR "computed tomography image"[All Fields] OR "computed tomography images"[All Fields] OR "computed tomographic scan"[All Fields] OR "computed tomographic scans"[All Fields]  OR "computed tomography scan"[All Fields] OR "computed tomography scans"[All Fields] OR "ultrasound"[All Fields] OR "mammographic images"[All Fields] OR "mammographic image"[All Fields] OR "mammogram"[All Fields] OR "mammograms"[All Fields] OR "mammography image"[All Fields] OR "mammography images"[All Fields] OR "skin lesion"[All Fields] OR "skin lesions"[All Fields] OR "Endoscopic images"[All Fields] OR "Endoscopic image"[All Fields] OR "Endoscopy image"[All Fields] OR "Endoscopy images"[All Fields] OR "radiograph"[All Fields] OR "radiographs"[All Fields] OR "radiographic image"[All Fields] OR "radiographic images"[All Fields] OR "radiography image"[All Fields] OR "radiography images"[All Fields] OR "x-ray"[All Fields] OR "x-rays"[All Fields] OR "fundus image"[All Fields] OR "fundus images"[All Fields] OR "optical coherence tomography image"[All Fields] OR  "optical coherence tomography images"[All Fields] OR "OCT image"[All Fields] OR "OCT images"[All Fields] OR "cephalogram"[All Fields] OR "cephalograms"[All Fields] OR "cephalometric image"[All Fields] OR "cephalometric images"[All Fields] OR "dermoscopic images"[All Fields] OR "dermoscopic image"[All Fields] OR "dermoscopy images"[All Fields] OR "dermoscopy image"[All Fields]) |
| **IEEE** | ("transfer learning" OR "deep learning" OR "convolutional neural network" OR "convolutional neural networks") AND ( "MRI" OR "MRIs" OR "Magnetic resonance images" OR "Magnetic resonance image" OR "MR image" OR "MR images" OR "CT" OR "CTs" OR "computed tomographic image" OR "computed tomographic images" OR "computed tomography image" OR "computed tomography images" OR "computed tomographic scan" OR "computed tomographic scans" OR "computed tomography scan" OR "computed tomography scans" OR "ultrasound" OR "mammographic images" OR "mammographic image" OR "mammogram" OR "mammograms" OR "mammography image" OR "mammography images" OR "skin lesion" OR "skin lesions" OR "endoscopic images" OR "endoscopic image" OR "endoscopy image" OR "endoscopy images" OR "radiograph" OR "radiographs" OR "radiographic image" OR "radiographic images" OR "radiography image" OR "radiography images" OR "x-ray" OR "x-rays" OR "fundus image" OR "fundus images" OR "optical coherence tomography image" OR  "optical coherence tomography images" OR "OCT image" OR "OCT images" OR "cephalogram" OR "cephalograms" OR "cephalometric image" OR "cephalometric images" OR "dermoscopic images" OR "dermoscopic image" OR "dermoscopy images" OR "dermoscopy image") |
| **ACM digital library** | ("transfer learning" OR "deep learning" OR "convolutional neural network" OR "convolutional neural networks") AND ( "MRI" OR "MRIs" OR "Magnetic resonance images" OR "Magnetic resonance image" OR "MR image" OR "MR images" OR "CT" OR "CTs" OR "computed tomographic image" OR "computed tomographic images" OR "computed tomography image" OR "computed tomography images" OR "computed tomographic scan" OR "computed tomographic scans" OR "computed tomography scan" OR "computed tomography scans" OR "ultrasound" OR "mammographic images" OR "mammographic image" OR "mammogram" OR "mammograms" OR "mammography image" OR "mammography images" OR "skin lesion" OR "skin lesions" OR "endoscopic images" OR "endoscopic image" OR "endoscopy image" OR "endoscopy images" OR "radiograph" OR "radiographs" OR "radiographic image" OR "radiographic images" OR "radiography image" OR "radiography images" OR "x-ray" OR "x-rays" OR "fundus image" OR "fundus images" OR "optical coherence tomography image" OR  "optical coherence tomography images" OR "OCT image" OR "OCT images" OR "cephalogram" OR "cephalograms" OR "cephalometric image" OR "cephalometric images" OR "dermoscopic images" OR "dermoscopic image" OR "dermoscopy images" OR "dermoscopy image") |

**Table S2**: List of abbreviations.

| Abbreviation | Full | Abbreviation | Full |
|---|---|---|---|
| TraT | Transformation Type | HM | Heatmap |
| ClassT | Classification Task | Ag | Augmentation |
| FinalC | Final Classifier | Acc | Accuracy |
| Vis | Visualization Method | Sen | Sensitivity |
| FE | Feature Extraction | RF | Random Forest |
| FC | Fully Connected Layer | DC | Deconvolution |
| SVM | Support Vector Machine | DiceC | Dice Coefficient |
| AM | Activation Maximization | FT | Fine-tuning |
| LDA | Linear Discriminant Analysis | AntS | Anatomical site |
| MRI | Magnetic resonance imaging | CT | Computed tomography |
| OCT | Optical coherence tomography | X-ray | X-radiation |

**Table S3**: CT scan image studies and the extracted features according to Table 1.

| Paper | AntS | Medical task | Method | Data Size | Ag Data Size | Performance | TraT | ClassT | FinalC | Benchmark | Vis |
|---|---|---|---|---|---|---|---|---|---|---|---|
| Dawud et al. 2019[9] | Brain | Brain haemorrhage classification | AlexNet | 2,104 | 12,635 | Acc=93.48% | FT | Binary | SVM | | DC |
| Peng et al. 2020[108] | Liver | Transarterial chemoembolization prediction | ResNet-50 | 1,687 | 8,435 | Acc>82.8% | FT | 4 classes | FC | | |
| Shin et al. 2016[125] | Lung | Interstitial lung disease classification | AlexNet | 905 | 10,860 | Acc=90.2% | FT | 6 classes | FC | GoogLeNet | AM |
| Da Nóbrega et al. 2018[109] | Lung | Lung nodule classification | ResNet-50 | 7,371 | | AUC=0.93 | FE | Binary | SVM | VGG-16, VGG-19, Inception-V3, Xception, Inception-ResNet-V2, DenseNet-169, DenseNet-201 | |
| Nishio et al. 2018[85] | Lung | Lung nodule classification | VGG-16 | 1,236 | | Acc=68.0% | FT | 3 classes | FC | | |
| Lee et al. 2019[107] | Thyroid | Cervical lymph node metastasis diagnosis | ResNet-50 | 995 | | Acc=90.4% | FE | Binary | FC | Inception-V3 | HM |
| Santin et al. 2019[87] | Thyroid | Abnormalities of thyroid cartilage detection | VGG-16 | 515 | 2,575 | AUC=0.72 | FT | Binary | FC | | |
| Chowdhury et al. 2019[61] | Osteomeatal complex | Osteomeatal complex inflammation classification | Inception-V3 | 956 | | Acc=85% | FE | Binary | FC | | |
| Kajikawa et al. 2018[126] | Prostate | Dosimetric eligibility prediction | AlexNet | 480 | | Acc=70% | FT | Binary | FC | | HM |
| Lee et al. 2019[60] | Tooth | Cystic lesions classification | Inception-V3 | 2,126 | 212,600 | AUC>0.847 | FT | Binary | FC | | |
| Belharbi et al. 2017[86] | Vertebra | Spotting L3 slice | VGG-16 | 642 | | MAE=1.91 | FT | Numeric | FC | GoogLeNet, VGG-19, AlexNet | |

**Table S4**: MRI scan image studies and the extracted features according to Table 1.

| Paper | AntS | Medical task | Method | Data Size | Ag Data Size | Performance | TraT | ClassT | FinalC | Benchmark | Vis |
|---|---|---|---|---|---|---|---|---|---|---|---|
| Langner et al. 2019[75] | Body | Morphological indicators of aging identification | VGG-16 | 23,905 | | MAE=2.49 | FE | Numeric | FC | | |
| Zhang et al. 2019[119] | Brain | Visual response prediction | AlexNet | 1,750 | | Acc=98.6% | FE | Binary | FC | | |
| Maqsood et al. 2019[120] | Brain | Alzheimer's disease stages classification | AlexNet | 382 | | Acc=92.85% | FE | 4 classes | FC | | DC |
| Wang et al. 2019[121] | Brain | Alcoholism classification | AlexNet | 379 | 48,320 | Acc=97.42 | FT | Binary | FC | | |
| Afzal et al. 2019[123] | Brain | Alzheimer's stage detection | AlexNet | 218 | 6,104 | Acc=98.44% | FT | Binary | FC | | |
| Deepak and Ameer 2019[47] | Brain | Brain tumor classification | GoogLeNet | 3,064 | | Acc=98% | FT | 3 classes | KNN | | |
| Yang et al. 2019[51] | Brain | Glioma grading | GoogLeNet | 113 | 1,582 | Acc=0.867 | FT | Binary | FC | AlexNet | |
| Talo et al. 2019[99] | Brain | Brain disease detection | ResNet-50 | 1,074 | | Acc=95.23% | FT | 5 classes | FC | AlexNet, VGG-16, ResNet-18, ResNet-34 | AM |
| Gao et al. 2019[100] | Brain | Behavior tasks decoding | ResNet-34 | 965 | | Acc=75.0% | FE | Binary | FC | Inception-V3, AlexNet | HM |
| Korfiatis et al. 2017[98] | Brain | Methylation of the O6-methylguanine methyltransferase (MGMT) gene status prediction | ResNet-50 | 10,468 | | Acc=94.9% | FE | 3 classes | FC | ResNet-18, ResNet-34 | DC |
| Swati et al. 2019[74] | Brain | Brain tumor detection | VGG-19 | 3,064 | | Prec=96.13 | FT | Binary | FC | | DC |
| Khan et al. 2019[76] | Brain | Alzheimer's disease diagnosis | VGG-19 | 3,200 | | Acc>92.0% | FT | Binary | FC | | HM |
| Zhu et al. 2019[50] | Breast | Occult invasive disease prediction | GoogLeNet | 131 | 30,426 | AUC=0.70 | FE | Binary | SVM | | |
| Zhu et al. 2019[49] | Breast | Radiogenomic associations in breast cancer detection | GoogLeNet | 275 | 44,660 | AUC=0.65 | FE | Binary | SVM | VGG-16 | |
| Dallora et al. 2019[48] | Knee | Age assessment | GoogLeNet | 402 | 2,010 | MAE=0.98 | FT | Numeric | FC | ResNet-50 | |
| Yuan et al. 2019[122] | Prostate | Prostate cancer classification | AlexNet | 221 | 4,641 | Acc=86.92% | FT | Binary | FC | | |
| Zhong et al. 2019[101] | Prostate | Prostate cancer classification | ResNet-50 | 169 | 5,154 | AUC=0.726 | FT | Binary | FC | | |

**Table S5**: Ultrasound image studies and the extracted features according to Table 1.

| Paper | AntS | Medical task | Method | Data Size | Ag Data Size | Performance | TraT | ClassT | FinalC | Benchmark | Vis |
|---|---|---|---|---|---|---|---|---|---|---|---|
| Cheng et al. 2016[83] | Abdomen | Abdominal ultrasound image classification | VGG-16 | 5,518 | | Acc=77.9% | FE | 11 classes | FC | CaffeNet | |
| Cao et al. 2019[134] | Breast | Breast lesion detection | DenseNet-161 | 1,043 | | Acc>80.0% | FE | Binary | FC | AlexNet, ZFNet, VGG-16, ResNet-50, GoogLeNet | |
| Xiao et al. 2018[54] | Breast | Breast masses classification | Inception-V3 | 2,058 | 6,174 | Acc=85.13% | FT | Binary | FC | ResNet-50, Xception | |
| Fujioka et al. 2019[59] | Breast | Breast mass lesion classification | GoogLeNet | 947 | | Acc=92.5% | FE | Binary | FC | | |
| Byra et al. 2019[82] | Breast | Breast mass classification | VGG-19 | 882 | 5,292 | AUC=0.936 | FT | Binary | FC | | |
| Kuo et al. 2019[106] | Kidney | Chronic kidney disease (CKD) prediction | ResNet-101 | 4,505 | 37,696 | Acc=85.6% | FT | Binary | FC | | |
| Xue et al. 2020[56] | Liver | Liver fibrosis grading | Inception-V3 | 2,330 | 6,990 | AUC=0.95 | FT | Binary | FC | | |
| Byra et al. 2018[137] | Liver | Liver steatosis assessment | Inception-ResNet-V2 | 550 | | AUC=0.977 | FE | Binary | SVM | | |
| Song et al. 2019[55] | Thyroid | Thyroid nodules diagnosis | Inception-V3 | 1,358 | | Sen=95.2% | FE | Binary | FC | | |
| Chi et al. 2017[57] | Thyroid | Thyroid nodule classification | GoogLeNet | 428 | 3,852 | Acc=98.2% | FT | Binary | FC | | |
| Guan et al. 2019[58] | Thyroid | Thyroid nodule classification | Inception-V3 | 2,836 | | Sen=93.3% | FT | Binary | FC | | |
| Qin et al. 2019[84] | Thyroid | Thyroid nodules classification | VGG-16 | 233 | 1,156 | Acc=86.21% | FE | Binary | FC | ResNet-18, GoogLeNet, Inception-V3, AlexNet | HM |

**Table S6**: Skin Lesion image studies and the extracted features according to Table 1.

| Paper | Medical task | Method | Data Size | Ag Data Size | Performance | TraT | ClassT | FinalC | Benchmark | Vis |
|---|---|---|---|---|---|---|---|---|---|---|
| Hosny et al. 2019[127] | Skin lesion classification | AlexNet | 206 | 14,832 | Acc>95.91% | FT | 3 classes | FC | | |
| Cui et al. 2019[66] | Melanoma diagnosis | Inception-V3 | 2,200 | | Acc=93.74% | FE | Binary | FC | AlexNet, VGG-16, VGG-19 | |
| Binol et al. 2019[138] | Rosacea identification | Inception-ResNet-V2 | 10,922 | Online | DiceC=89.8% | FT | Binary | FC | ResNet-101 | |
| Kassani et al. 2019[112] | Melanoma detection | ResNet-50 | 9,887 | 34,577 | Acc=92.0% | FT | 7 classes | FC | Xception, VGG-16, VGG-19 | DC |
| Lopez et al. 2017[91] | Skin lesion classification | VGG-16 | 1,279 | 7,782 | Acc=81.3% | FT | Binary | FC | | |
| Kwasigroch et al. 2017 [90] | Skin lesion classification | VGG-19 | 1,803 | 6,498 | Acc=80.7% | FT | Binary | FC | ResNet-50 | |
| Yu et al. 2018[89] | Melanoma detection | VGG-16 | 724 | 940 | Acc=83.5% | FT | Binary | FC | | DC |

**Table S7**: Fundus image studies and the extracted features according to Table 1.

| Paper | Medical task | Method | Data Size | Ag Data Size | Performance | TraT | ClassT | FinalC | Benchmark | Vis |
|---|---|---|---|---|---|---|---|---|---|---|
| Li et al. 2016[124] | Glaucoma diagnosis | AlexNet | 650 | | AUC=0.83 | FE | Binary | FC | VGG-19, VGG-16, GoogLeNet | |
| Li et al. 2019[52] | Diabetic retinopathy detection | Inception-V3 | 8,816 | | Acc=93.49% | FT | 5 classes | FC | | |
| Arcadu et al. 2019[53] | Optical coherence tomography measures detection | Inception-V3 | 30,371 | | AUC=0.97 | FT | Binary | FC | | |
| Lu et al. 2019[102] | Optic disc laterality detection | ResNet-152 | 576 | | Acc=97.2% | FT | Binary | FC | | |
| Hemelings et al. 2019[103] | Glaucoma detection | ResNet-50 | 1,775 | 7,038 | AUC=0.995 | FT | Binary | FC | | HM |
| Christopher et al. 2018[104] | Glaucomatous optic neuropathy identification | ResNet-50 | 14,822 | 148,220 | AUC=0.91 | FT | Binary | FC | VGG-16, Inception-V3 | HM |
| Li et al. 2019[105] | Cataract diagnosis | ResNet-50 | 8,030 | | Acc=87.7% | FE | 4 classes | FC | ResNet-18 | HM |
| Gómez-Valverde et al. 2019[77] | Glaucoma detection | VGG-19 | 2,313 | Online | AUC=0.94 | FT | Binary | FC | GoogLeNet, ResNet-50 | |
| Choi et al. 2017[78] | Retinal disease detection | VGG-19 | 279 | 10,000 | Acc=72.8% | FE | 10 classes | RF | AlexNet | |
| Li et al. 2018[79] | Diabetic retinopathy classification | VGG-19 | 1,014 | 15,210 | Acc>92.01% | FT | 4 classes | FC | AlexNet, GoogLeNet, VGG-16 | |
| Zhang et al. 2019[80] | Retinopathy of prematurity screening | VGG-16 | 382,922 | | Acc=99.88% | FT | Binary | FC | AlexNet, GoogLeNet | |
| Nagasato et al. 2019[81] | Branch retinal vein detection | VGG-16 | 466 | 8,388 | AUC=0.97 | FT | Binary | FC | | |

**Table S8**: OCT image studies and the extracted features according to Table 1.

| Paper | Medical task | Method | Data Size | Ag Data Size | Performance | TraT | ClassT | FinalC | Benchmark | Vis |
|---|---|---|---|---|---|---|---|---|---|---|
| Islam et al. 2019[135] | Diabetic retinopathy identification | DenseNet-201 | 109,309 | | Acc=97.0% | FT | 4 classes | FC | AlexNet, VGG-16, ResNet-18, VGG-19, GoogLeNet, Inception-V3, ResNet-50, ResNet-101, Inception-ResNet-V2 | |
| Kermany et al. 2018[67] | Diabetic retinopathy classification | Inception-V3 | 207,130 | | Acc=96.6% | FE | 4 classes | FC | | HM |
| Lu et al. 2018[113] | Diabetic retinopathy diagnosis | ResNet-101 | 25,134 | | Acc>84.8% | FE | Binary | FC | | |
| An et al. 2019[92] | Glaucoma diagnosis | VGG-19 | 347 | 1,041 | AUC>0.94 | FT | Binary | RF | | HM |
| Feng et al. 2019[93] | Retinal disorders detection | VGG-16 | 109,312 | | Acc=98.6% | FT | 4 classes | FC | | |
| Kaveri et al. 2019[94] | Glaucoma detection | VGG-16 | 737 | | AUC>0.93 | FE | Binary | RF | Inception-V3, ResNet-18 | HM |

**Table S9**: X-Ray image studies and the extracted features according to Table 1.

| Paper | AntS | Medical task | Method | Data Size | Ag Data Size | Performance | TraT | ClassT | FinalC | Benchmark | Vis |
|---|---|---|---|---|---|---|---|---|---|---|---|
| Huynh et al. 2016[114] | Breast | Mammographic tumor classification | AlexNet | 607 | | AUC=0.81 | FE | Binary | SVM | | |
| Zhang et al. 2018[115] | Breast | Mammogram and tomosynthesis image classification | AlexNet | 3,290 | 26,320 | AUC=0.72 | FT | Binary | FC | | |
| Li et al. 2017[116] | Breast | Breast cancer risk assessment | AlexNet | 456 | | AUC=0.82 | FE | Binary | SVM | | |
| Ragab et al. 2019[118] | Breast | Breast cancer detection | AlexNet | 1,318 | 5,272 | Acc=87.2% | FT | Binary | SVM | | |
| Jiang et al. 2017[43] | Breast | Breast mass lesion classification | GoogLeNet | 736 | 2,944 | AUC=0.88 | FT | Binary | FC | AlexNet | |
| Mednikov et al. 2018[44] | Breast | Breast cancer detection | Inception-V3 | 410 | 100,000 | AUC=0.91 | FT | Binary | FC | | |
| Arefan et al. 2020[45] | Breast | Breast cancer risk prediction | GoogLeNet | 678 | | AUC=0.73 | FT | Binary | LDA | | HM |
| Yi et al. 2019[95] | Breast | Breast mass lesion classification | ResNet-50 | 3,034 | Online | AUC=0.93 | FT | Binary | FC | | HM |
| Pan et al. 2019[132] | Chest | Abnormality detection in chest radiographs | DenseNet-121 | 17,202 | Online | AUC=0.90 | FT | Binary | FC | | |
| Dunnmon et al. 2019[133] | Chest | Abnormality detection in chest radiographs | DenseNet-121 | 216, 431 | | Acc=0.91 | FE | Binary | FC | AlexNet, ResNet-18 | HM |
| Rajkomar et al. 2017[46] | Chest | Abnormality detection in chest radiographs | GoogLeNet | 1,505 | 159,530 | Acc=99.7% | FT | Binary | FC | | |
| Zhou et al. 2019[27] | Heart | Cardiomegaly classification | Inception-V3 | 108,948 | | AUC=0.86 | FE | 8 classes | FC | ResNet-50, Xception | DC |
| Yu et al. 2019[41] | Hip | Hip fracture detection | Inception-V3 | 617 | | Acc>90.9% | FE | 4 classes | FC | | HM |
| Abidin et al. 2018[39] | Knee | Chondrocyte patterns classification | Inception-V3 | 842 | | AUC>0.95 | FE | Binary | SVM | CaffeNet | |
| Yi et al. 2019[96] | Knee | Knee arthroplasty classification | ResNet-18 | 158 | 1,274 | AUC=1.0 | FT | Binary | FC | | HM |
| Abbas et al. 2018[117] | Lung | Manifestation of tuberculosis identification | AlexNet | 138 | 60,000 | AUC=0.99 | FT | Binary | FC | | DC |
| Gozes et al. 2019[128] | Lung | Tuberculosis detection | DenseNet-121 | 112,000 | Online | AUC=0.965 | FT | Binary | FC | | |
| Nguyen et al. 2019[130] | Lung | Tuberculosis detection | DenseNet-121 | 18,686 | 112,120 | AUC=0.89 | FT | 14 classes | FC | VGG-16, VGG-19, ResNet-50, Inception-ResNet-V2 | HM |
| Varshni et al. 2019 [131] | Lung | Pneumonia detection | DenseNet-169 | 2,862 | | AUC=0.80 | FE | Binary | SVM | VGG-16, VGG-19, ResNet-50, Xception | |

**Table S9 (Continued)**: X-Ray image studies and the extracted features according to Table 1.

| Paper | AntS | Medical task | Method | Data Size | Ag Data Size | Performance | TraT | ClassT | FinalC | Benchmark | Vis |
|---|---|---|---|---|---|---|---|---|---|---|---|
| Ahsan et al. 2019[69] | Lung | Tuberculosis detection | VGG-16 | 1,324 | Online | Acc>78.3% | FT | Binary | FC | | HM |
| Yi et al. 2019[97] | Skeletal system | Pediatric musculoskeletal radiographs classification | ResNet-18 | 250 | 7,500 | AUC=1.0 | FT | 5 classes | FC | | HM |
| HJ et al. 2020[129] | Tooth | Skeletal classification | DenseNet-121 | 5,890 | 50,000 | Acc>90% | FT | 3 classes | FC | | HM |
| Lee et al. 2018[42] | Tooth | Dental caries diagnosis | Inception-V3 | 3,000 | 30,000 | Acc>82.0% | FT | 4 classes | FC | | |
| Poedjiastoeti et al. 2018[68] | Tooth | Jaw tumor diagnosis | VGG-16 | 500 | 1,000 | Acc=83.0% | FT | Binary | FC | | HM |
| Lee et al. 2020[72] | Tooth | Osteoporosis in dental panoramic radiographs classification | VGG-16 | 680 | | Acc=84.0% | FT | Binary | FC | | HM |
| Prajapati et al. 2017[70] | Tooth | Dental diseases classification | VGG-16 | 250 | | Acc=88.5% | FE | 3 classes | FC | | |
| Lee et al. 2018[73] | Tooth | Periodontally compromised teeth diagnosis | VGG-19 | 1,740 | 104,400 | Acc>76.7% | FT | Binary | FC | | |
| Kim et al. 2018[40] | Wrist | Fracture detection | Inception-V3 | 1,389 | 11,112 | AUC=0.954 | FT | Binary | FC | | |
| Han et al. 2018[136] | Wrist | Bone age assessment | Inception-ResNet-V2 | 12,611 | | MAE=15.16 | FE | Numeric | SVR | VGG-16, VGG-19, ResNet-50, Inception-V3, Xception | |
| Yune et al. 2019[71] | Wrist | Gender classification | VGG-16 | 10,318 | | Acc=95.9% | FT | Binary | FC | | HM |

**Table S10**: Endoscopy image studies and the extracted features according to Table 1.

| Paper | AntS | Medical task | Method | Data Size | Ag Data Size | Performance | TraT | ClassT | FinalC | Benchmark | Vis |
|---|---|---|---|---|---|---|---|---|---|---|---|
| Wimmer et al. 2017[88] | Stomach | Celiac disease diagnosis | VGG-16 | 1,661 | Online | Acc=90.5% | FT | Binary | FC | AlexNet | |
| Liu et al. 2018[62] | Stomach | Gastric cancer diagnosis | Inception-V3 | 2,331 | 16,317 | Acc=98.5% | FT | Binary | FC | VGG-16, Inception-ResNet-V2 | AM |
| Sakai et al. 2018[64] | Stomach | Gastric cancer diagnosis | GoogLeNet | 29,037 | 348,943 | AUC=0.95 | FT | Binary | FC | | |
| Li et al. 2020[65] | Stomach | Early gastric cancer diagnosis | Inception-V3 | 2,088 | 20,000 | Acc=90.9% | FT | Binary | FC | | |
| Lee et al. 2019[110] | Stomach | Gastric cancer diagnosis | ResNet-50 | 787 | | AUC=0.97 | FT | 3 classes | FC | Inception-V3, VGG-16 | |
| Zhu et al. 2019[111] | Stomach | Invasion depth of gastric cancer diagnosis | ResNet-50 | 993 | | Acc=89.16% | FE | Binary | FC | | |
| Li et al. 2017[63] | Gastrointestinal tract | Gastrointestinal bleeding detection | Inception-V3 | 2,890 | 5,410 | Acc=98.62% | FE | Binary | FC | | |